\documentclass[a4paper,11pt,draft]{article}

\usepackage{latexsym}
\usepackage{amssymb}
\usepackage{amsfonts}
\usepackage{amsmath}

\newcommand{\bol}[1]{\boldsymbol {#1}}
\newcommand{\sss}[1]{\scriptscriptstyle {#1}}

\def\d{\partial}
\def\f{\phi}

\begin{document}

\author{Nicola Grillo \\[3mm]
{\textit{Institut f\"ur Theoretische Physik der Universit\"at Z\"urich}} \\
{\textit{Winterthurerstrasse 190, CH-8057 Z\"urich}}\\[4mm]
{\texttt{grillo@physik.unizh.ch}} }

\title{Causal Quantum Gravity}

\maketitle

\begin{abstract}\noindent
I discuss some issues of perturbative quantum gravity, namely of  
a theory  of self-interacting massless spin-2 quantum gauge fields, 
the \textit{gravitons}, on flat space-time, in the framework of causal 
perturbation theory. The central aspects of this approach  lie in  the 
construction of the  scattering matrix by means of causality and Poincar\'e 
covariance and in the analysis of the gauge structure of the theory.
For this  purpose, two main tools will be used:  the Epstein--Glaser 
inductive and causal construction of the perturbation series for the 
scattering matrix  and  the concept of perturbative operator quantum 
gauge invariance borrowed from non-Abelian quantum gauge theories.
The first method deals with the ultraviolet problem of quantum gravity  
and the second one ensures gauge invariance at the quantum level, 
formulated by means of a gauge charge, in each order of perturbation theory.
The gauge charge leads to a characterization of the physical subspace 
of the graviton Fock space. Aspects of quantum gravity coupled to scalar 
matter fields are also discussed.
\end{abstract}

\vfill

\begin{center}

{\large\textsl{Second International School on Field Theory and Gravitation}} 

{\large\textsl{Vit\'oria -- Brazil, April 25 -- 28, 2000}}

\end{center}
\newpage
\tableofcontents
\newpage
\section{Introduction}
\setcounter{equation}{0}

The central aspect of this work is the construction of the $S$-matrix for gravity
by means of causality in the quantum field theoretical (QFT) framework.
This idea goes back to St\"uckelberg, Bogoliubov and Shirkov and the program
was carried out successfully by Epstein and Glaser~\cite{eg,piguet}  for scalar field theories and
subsequently applied to QED by Scharf~\cite{Scha1}, to non-Abelian gauge theories
by D\"utsch {\it et al.}~\cite{ym1,ym2,Aste1} and  to quantum gravity (QG),
(by which we mean  a QFT of \textit{self-interacting massless spin-2 quantum gauge fields on
flat space-time}), by Schorn~\cite{schorn1,schorn2}. For this non-geometrical approach, 
see ~\cite{Feynman,ogiepolu,wyss}.
For our purpose, namely the implementation of QG as a Poincar\'e covariant local  quantum field theory
with a considerable gauge arbitrariness, two main tools will be used: 
the Epstein-Glaser inductive construction
of the perturbation series for the $S$-matrix with the related causal renormalization
scheme~\cite{eg,Scha1} and  the concept of perturbative operator quantum gauge invariance~\cite{Aste1,Schawell}.
A detailed exposition of what follows can be found in~\cite{gri3,gri4,gri5,gri6}.

\section{Causal Perturbation Theory}
\label{sec:smatrix}
\setcounter{equation}{0}

In this section we give a concise review of the causal approach to QFT.
We consider the $S$-matrix, being a formal power series in the coupling constant,
as a sum of smeared operator-valued distributions of the following form~\cite{eg,piguet,Scha1}
\begin{equation}
S(g)={\mathbf 1}+\sum_{n=1}^{\infty}\frac{1}{n!} \int d^{4}x_{1}
\ldots d^{4}x_{n}\, T_{n}(x_{1},\ldots,
x_{n})\,g(x_{1})\cdot\ldots\cdot g(x_{n})\, .
\label{eq:1}
\end{equation}
The  Schwartz test function $g\in\mathcal{S}(\mathbb{R}^{4})$ plays the r{\^o}le of
adiabatically  switching  the interaction and provides a natural infrared cutoff
in the long-range part of the interaction.

To establish the  existence of the \textit{adiabatic limit} $g\to 1$ in theories 
involving self-coupled massless particles, like QG, may be problematic. This aspect will not
be considered here.

The $n$-point operator-valued distributions $T_{n}$
are well-defined  \textit{renormalized time-ordered products} and can be
expressed in terms of Wick monomials of free fields. They are constructed
inductively from the first order $T_{1}(x)$, which plays the r{\^o}le of
the usual interaction Lagrangian in terms of free fields, 
by means of Poincar\'e covariance and
causality. The latter, if correctly incorporated, leads directly to the renormalized
perturbation series for the $S$-matrix which is UV-finite in every order. 

The construction of
$T_{n}$ requires some care: if it were simply given by the usual time-ordering
\begin{equation}
\begin{split}
T_{n}(x_{1},\ldots,x_{n})&=\mathcal{T}\big\{T_{1}(x_{1})\ldots T_{1}(x_{n})\big\}\\
&=\sum_{\pi \in \sigma_{n}}\Theta (x^{0}_{\pi(1)}-x^{0}_{\pi(2)})\ldots\Theta(x^{0}_{\pi(n-1)}-x^{0}_{\pi(n)}) \\
&\quad\qquad\qquad\times  T_{1}(x_{\pi(1)})\ldots T_{1}(x_{\pi(n)})\, ,
\end{split}
\label{eq:2}
\end{equation}
then UV-divergences would appear. If the arguments
$x_{1},\ldots ,x_{n}$ are all time-ordered, {\em i.e.} if we have $x_{1}^{0}>x_{2}^{0}>
\ldots > x_{n}^{0}$, then $T_{n}$ is rigorously given by 
$T_{n}(x_{1},\ldots,x_{n})=T_{1}(x_{1})\ldots T_{1}(x_{n})$. Since $T_{n}$
has to be  totally symmetric in $x_{1},\ldots ,x_{n}$, we so obtain $T_{n}$
everywhere except for the complete diagonal $\Delta_{n}=\{x_{1}=x_{2}=\ldots =x_{n}\}$,
{\em i.e.} except for the coincident point in configuration space. 
The correct treatment of this point constitutes the key  to control 
the UV-behaviour of the $n$-point distributions. Indeed, products of Feynman propagators 
with coincident arguments
\begin{equation}
\begin{split}
\Pi(x-y)=&D_{m}^{\sss F}(x-y)\cdot D_{m}^{\sss F}(x-y)=\,?\\
\hat{\Pi}(p)\sim&\int\!\!d^{4}k\, \frac{1}{(p-k)^{2}-m^2 +i0}\,\frac{1}{k^2-m^2+i0}\\
            =&\text{ logarithmic divergent}
\end{split}
\end{equation}
are the origin of the UV-divergences in loop graphs, 
because time-ordering cannot be done simply by multiplying (singular) distributions 
by discontinuous $\Theta$-distributions, since this procedure is usually ill-defined.

The distributions must be carefully split into a retarded and an advanced part 
for the $T_{n}$ to be well-defined and finite.

Let us illustrate how the inductive construction  works by means of an example
in which $T_{2}(x_{1},x_{2})$ is constructed for a massive scalar field $\varphi$ 
with a $\varphi^{3}$-coupling.

We define a QFT by giving the equation of 
motion of the free quantum field, the covariant commutator rule and the interaction 
Lagrangian $T_{1}$ with coupling strength $g$
\begin{equation}
\left(  \Box + m^2 \right)\varphi(x)=0\, ,\quad \left[ \varphi(x_{1}),\varphi(x_{2})\right]=
-iD_{{m}}(x_{1}-x_{2})\, ,
\quad T_{1}(x)=i\,g\,:\! \varphi(x)^3\! :\, ,
\label{eq:3}
\end{equation}
where $D_{m}$ is the massive Jordan--Pauli distribution
\begin{equation}
D_{m}(x)=\frac{i}{(2\pi)^3}\int\!\! d^{4}p \, \delta(p^2 -m^2)\, \textrm{sgn}(p^0)\, e^{-i p\cdot x}\,  .
\label{eq:4}
\end{equation}
Causality of the $S$-matrix means
\begin{equation}
S(g_{1}+g_{2})=S(g_{2}) S(g_{1})\quad\text{for}\quad \mathrm{supp}(g_{1})< \mathrm{supp}(g_{2})\,.
\end{equation}
The notation $<$ in the support condition means more precisely:  
$\mathrm{supp}(g_{1})\cap\big\{ \mathrm{supp}(g_{2})+V^{\sss +}\big\}=\emptyset$.

Translated in terms of $T_{2}(x_{1},x_{2})$, the condition of causality  becomes
\begin{equation}
T_{2}(x_{1},x_{2})=\begin{cases}T_{1}(x_{1})T_{1}(x_{2})& \text{for}\;x_{1} > x_{2}\,,\\
                                T_{1}(x_{2})T_{1}(x_{1})& \text{for}\:x_{2} >  x_{1}\,.
\end{cases}
\end{equation}
Clearly, difficulties arise for $x_{1}=x_{2}$.

Following the causal construction of Epstein and Glaser, we define the auxi\-liary distributions
\begin{gather}
R_{2}'(x_{1},x_{2}):=  - T_{1}(x_{2})T_{1}(x_{1})\,,\quad A_{2}'(x_{1},x_{2}):= - T_{1}(x_{1})T_{1}(x_{2})\nonumber\\
D_{2}(x_{1},x_{2}):= R_{2}'(x_{1},x_{2})-A_{2}'(x_{1},x_{2})\, ,
\label{eq:5}
\end{gather}
and carry out  all  possible contractions in $D_{2}$ using Wick's lemma, leading to
\begin{equation}
D_{2}(x_{1},x_{2})=\sum_{k=1}^{3} :{\mathcal O}_{k}(x_{1},x_{2}):d_{2}^{\sss {\left[ k \right]}}(x_{1}-x_{2})\, .
\label{eq:6}
\end{equation}
$:{\mathcal O}_{k}(x_{1},x_{2}):$ represents a normally ordered product  of free field operators
and $d_{2}^{\sss{ \left[ k\right]}} (x_{1}-x_{2}) $ is a  numerical distribution.
Expanding the result we can identify tree, loop and vacuum graph contributions (no tadpoles appear), respectively
\begin{equation}
\begin{split}
D_{2}(x_{1},x_{2})=&+ :\!\varphi(x_{1})\varphi(x_{1})\varphi(x_{2})\varphi(x_{2})\!:d_{2}^{\sss
 {\left[1 \right]}}(x_{1}-x_{2})+\\
&+:\!\varphi(x_{1})\varphi(x_{2})\!:d_{2}^{\sss{\left[2 \right]}}(x_{1}-x_{2})+d_{2}^{\sss {\left[ 3 \right]}}(x_{1}-x_{2})\, ,
\label{eq:bb}
\end{split}
\end{equation}
where the numerical distributions are given by
\begin{align}
d_{2}^{\sss{\left[1 \right]}}(x_{1}-x_{2}) & =9ig^2 \big[ D_{{m}}^{\sss(+)}(x_{1}-x_{2})   
+D_{{m}}^{\sss(-)}(x_{1}-x_{2})  \big]=9ig^2 D_{{m}}(x_{1}-x_{2})\, ,\nonumber \\
d_{2}^{\sss{\left[2 \right]}}(x_{1}-x_{2}) & =18g^2 \big[ D_{{m}}^{\sss(+)}(x_{1}-x_{2})^2 -
D_{{m}}^{\sss(-)}(x_{1}-x_{2})^2\big] \, ,\nonumber \\
d_{2}^{\sss{\left[3 \right]}}(x_{1}-x_{2}) & =-6ig^2\big[ D_{{m}}^{\sss(+)}(x_{1}-x_{2})^3 
+D_{{m}}^{\sss(-)}(x_{1}-x_{2})^3\big] \, .
\end{align}
In addition, we define 
\begin{equation}
\begin{split}
R_{2}(x_{1},x_{2}):=& - T_{1}(x_{2})T_{1}(x_{1})+T_{2}(x_{1},x_{2})\,,\\
A_{2}(x_{1},x_{2}):=& - T_{1}(x_{1})T_{1}(x_{2})+T_{2}(x_{1},x_{2})\,;
\end{split}
\end{equation}
so that
\begin{equation}
D_{2}(x_{1},x_{2})= R_{2}(x_{1},x_{2})-A_{2}(x_{1},x_{2})\, .
\end{equation}
From this last equation, it follows that
\begin{equation}
T_{2}(x_{1},x_{2})= R_{2}(x_{1},x_{2})-R'_{2}(x_{1},x_{2})     \, .
\end{equation}
Now, the issue is how to compute $R_{2}$ without using its definition (since it contains the unknown $T_{2}$). 
This can be done
by analyzing the support property of $D_{2}$: 
the most important property of $D_{2}$ is causality, {\em i.e.} $\hbox{supp}(d_{2}^{\sss{\left[j \right]}} (x))
\subseteq \overline{ V^{+}(x)}
\cup \overline{ V^{-}(x)}$, with $x:=x_{1}-x_{2}$.

In order to obtain $T_{2}(x_{1},x_{2})$ we have to split the  distribution $D_{2}$ into a retarded part, $R_{2}$, and
an advanced part, $A_{2}$, with respect to the coincident point $x=0$, so that $\hbox{supp}(R_{2}(x))\subseteq \overline
{ V^{+}(x)}$ and $\hbox{supp}(A_{2}(x))\subseteq \overline{ V^{-}(x)}$.

This splitting of the numerical distribution $d_{2}^{\sss {\left[ k \right]}}(x)$ must be accomplished
 according to the correct singular order $\omega(d_{2}^{\sss {\left[ k \right]}})$, which agrees here
with the usual power-counting degree of Feynman diagrams, and describes the behaviour of 
$d_{2}^{\sss{\left[ k \right]}}(x)$ near $x=0$, or that of $\hat{d}_{2}^{\sss
 {\left[ k \right]}}(p)$ in the limit $p\to\infty$.

If $\omega < 0$, then the splitting is trivial and agrees with the standard time-ordering and we recover 
the Feynman rules.
If $\omega\ge 0$, then the splitting is non-trivial and non-unique
\begin{equation}
d_{2}^{\sss{\left[ k \right]}}(x)\longrightarrow r_{2}^{\sss {\left[ k \right]}}(x)+
\sum_{|a| =0}^{\omega(d_{2}^{\left[ k\right]})} C_{a}\, D^{a}\, \delta^{\sss (4)}(x)\, ,
\label{eq:7}
\end{equation}
and the retarded part $r_{2}^{\sss {\left[ k \right]}}(x)$ is  obtained in momentum space by
means of  a subtracted dispersion integral (thus recovering the relation between causality and dispersion relation)
of the form
\begin{equation}
\hat{r}_{2}^{\sss{\left[ k \right]}}(p)=\frac{i}{2\pi}\int_{-\infty}^{\infty}\! dt \frac{\hat{d}_{2}^{\sss
 {\left[ k \right]}}(tp)}{\left( t-i0\right)^{\omega +1}\left( 1-t+i0\right)}\quad ,\, p\in V^{+}\, .
\label{eq:8}
\end{equation}
Eq.~(\ref{eq:7}) contains a local ambiguity in the normalization: the $C_{a}$'s are undetermined finite 
normalization constants, which multiply terms with local support $\sim\delta^{\sss (4)}(x_{1}-x_{2})$.
This freedom in the normalization has to be restricted by further physical conditions, {\em e.g.} unitarity,
Lorentz covariance, existence of the adiabatic limit and  gauge invariance 
in the case of gauge theories or gravity.

Finally, $T_{2}$ is given by
\begin{equation}
T_{2}(x_{1},x_{2})=R_{2}(x_{1},x_{2})+T_{1}(x_{2})T_{1}(x_{1})=\sum_{k=0}^{3} :{\mathcal O}_{k}(x_{1},x_{2}):
t_{2}^{\sss {\left[ k \right]}}(x_{1}-x_{2})^{tot} \,  ,
\label{eq:9}
\end{equation}
with
\begin{equation}
\hat{t}_{2}^{\sss {\left[ k \right]}}(p)^{ tot}=\hat{t}_{2}^{\sss
 {\left[ k \right]}}(p)+\sum_{|a| =0}^{\omega(d_{2}^{\left[ k\right]})}\tilde{ C}_{a}\, p^a \, .
\label{eq:10}
\end{equation}
Applying the described scheme to our example, we find for the operator-valued distribution $T_{2}(x_{1},x_{2})$, Eq.~(\ref{eq:9}),
the expression
\begin{equation}
\begin{split}
T_{2}(x_{1},x_{2})& =  + :\!\varphi(x_{1})\varphi(x_{1})\varphi(x_{1})\varphi(x_{2})\varphi(x_{2})
\varphi(x_{2})\!:(-g^2)  \\
                            &\quad  + :\!\varphi(x_{1})\varphi(x_{1})\varphi(x_{2})\varphi(x_{2})\!:
                                                t_{2}^{\sss{\left[1 \right]}}(x_{1}-x_{2})^{tot}   \\
                  &\quad  + :\!\varphi(x_{1})\varphi(x_{2})\!:t_{2}^{\sss{\left[2 \right]}}(x_{1}-x_{2})^{tot} +
                                   t_{2}^{\sss{\left[3 \right]}}(x_{1}-x_{2})^{tot} \, .
\end{split}
\end{equation}
The first term represents the disconnected contribution coming, in Eq.~(\ref{eq:9}), from $T_{1}(x_{2})T_{1}(x_{1})$.
The distribution in the second term
$t_{2}^{\sss{\left[1 \right]}}(x_{1}-x_{2})^{tot}=+9\,i\,g^2\, D_{ m}^{\sss F}(x_{1}-x_{2})$ is the Feynman propagator
for the tree graph contribution, whereas the loop distribution $t_{2}^{\sss{\left[2 \right]}}(x_{1}-x_{2})^{tot}$ 
is easily obtained in momentum space by means of Eq.~(\ref{eq:8}) and reads
\begin{equation}
\hat{t}_{2}^{\sss{\left[2 \right]}}(p)^{tot}=\frac{i}{2\pi}\,\frac{-9g^2\pi}{(2\pi)^4}\,
    \int_{q}^{\infty}\!\! ds \frac{\sqrt{s(s-q)}}{s^2(1-s+i0)} +c_{0}\ , 
\quad c_{0}\in \mathbb{R}\, ,\  q=\frac{4m^2}{p^2}\, .
\end{equation}
The result of the  evaluation of the above integral can be found in Sec.~\ref{sec:matter}. 
Since $\omega(d_{2}^{\sss{\left[2 \right]}})=0$, the splitting is not unique and we must take the local normalization term $c_{0}$
into account. 

We do not give here the expression for the vacuum graph contribution, the treatment of the latter can be found in
Sec.~\ref{sec:causal} and in Sec.~\ref{sec:matter}.

The inductive construction can be repeated for every order of perturbation theory, although the complexity
increases. The most delicate step is the distribution splitting, which corresponds to a \textit{natural} and 
mathematical well defined ultraviolet regularization in the usual terminology.
The advantage of the causal scheme is that it leads directly to the  renormalized perturbative
expansion for the $S$-matrix without using a cutoff.  
It makes possible to compute finite amplitudes for various processes to a given order in the coupling 
constant   and it does not rely on the Lagrangian approach.

\section{Quantization of Gravity}
\label{sec:gravity}
\setcounter{equation}{0}

\subsection{From General Relativity to Quantum Gravity}

Since we are interested in a quantum theory of Einstein's general relativity,
we  start from the Hilbert--Einstein Lagrangian density
${\mathcal{L}}_{\sss HE}$ written in terms of the Goldberg
variable $\tilde{g}^{\mu\nu}=\sqrt{-g}\,g^{\mu\nu}$ and we  expand it into
a power series in the coupling constant $\kappa^{2} =32\,\pi\, G $, by introducing
the \textit{graviton} field $h^{\mu\nu}$ defined through
$\kappa\, h^{\mu\nu}=\tilde{g}^{\mu\nu}-\eta^{\mu\nu}$, where $\eta^{\mu\nu}=\mathrm{diag}(1,-1,-1,-1)$ is the 
flat space-time metric tensor
\begin{equation}
{\mathcal{L}}_{\sss HE}={\frac{- 2}{\kappa ^ 2}}\,\sqrt{-g}\, g^{\mu\nu}\,R_{\mu\nu} =
\sum_{j=0}^{\infty}\kappa^{j}\, {\mathcal{L}}_{\sss HE}^{\sss (j)}\, .
\label{eq:11}
\end{equation}
${\mathcal{L}}_{ \sss HE}^{\sss (j)}$ represents  an \textit{interaction}
involving $j+2$ gravitons.
From this formulation of general relativity we extract the ingredients
for the perturbative construction of causal QG. 

We stress however the
fact that we consider the classical Lagrangian density Eq.~(\ref{eq:11})
only as a \textit{source} of information about the fields, the couplings and the
gauge which we work with.  Causal perturbation theory does not  rely on a
quantum Lagrangian with interacting fields. 

In a  new approach, which  has been proposed in~ \cite{Schawell}, 
one constructs
the first-order interaction essentially by the requirement of perturbative quantum gauge 
invariance (see Sec.~\ref{sec:struttura}).

By considering the Euler--Lagrange variation of ${\mathcal{L}}_{\sss HE}^{\sss (0)}$
from  Eq.~(\ref{eq:11}) in the Hilbert gauge $h^{\alpha\beta}(x)_{,\beta}=0$ we
obtain the equation of motion for the free graviton field
$\Box h^{\alpha\beta}(x)=0$.

\subsection{Quantum Gravity as a Quantum Field Theory}

In  quantum gravity, we consider the  free rank-2 quantum tensor field $h^{\mu\nu}(x)$, the \textit{graviton},
which fulfils the free wave equation after having fixed the gauge. For the causal construction we need
the commutation relation between free field  operators
at different space-time points and the first-order graviton self-coupling $T_{1}^{h}(x)$.

The graviton field fulfils the Lorentz covariant  quantization rule
\begin{equation}
\begin{split}
\left[ h^{\alpha\beta}(x),h^{\mu\nu}(y) \right]=&-
\frac{i}{2}\,\left(  \eta^{\alpha\mu}\eta^{\beta\nu}+ 
  \eta^{\alpha\nu}\eta^{\beta\mu}-\eta^{\alpha\beta}\eta^{\mu\nu}\right)\, D_{0}(x-y)\\
=&\, -i\,  b^{\alpha\beta\mu\nu}\,D_{0}(x-y)\, ,
\label{eq:12}
\end{split}
\end{equation}
where $D_{0}(x)$ is the massless  Jordan--Pauli causal distribution.

The first order coupling among gravitons, being linear in the coupling constant $\kappa$,
can be derived from Eq.~(\ref{eq:11}) by taking the normally ordered product of ${\mathcal{L}}_{\sss HE}^
{\sss (1)}$ 
\begin{equation}
\begin{split}
T_{1}^{h,\sss HE}(x)& =i\,\kappa :{\mathcal{L}}_{\sss HE}^{\sss (1)}(x):=i\, \frac{\kappa}{2}\,  \big\{
            :h^{\rho\sigma}(x)h^{\alpha\beta}(x)_{,\rho}h^{\alpha\beta}(x)_{,\sigma}:+\\
&\quad-\frac{1}{2}:h^{\rho\sigma}(x)h(x)_{,\rho}h(x)_{,\sigma}:  
         +2:h^{\rho\sigma}(x)h^{\sigma\beta}(x)_{,\alpha}h^{\rho\alpha}(x)_{,\beta}:\\
            &\quad   +:h^{\rho\sigma}(x)h(x)_{,\alpha}h^{\rho\sigma}(x)_{,\alpha}:
          -2:h^{\rho\sigma}(x)h^{\alpha\rho}(x)_{,\beta}h^{\sigma\alpha}(x)_{,\beta}: \big\}\, .
\label{eq:13}
\end{split}
\end{equation}
The non-linearity of gravitation reflects itself in the self-coupling of gravitons.
For convenience of notation, $h:=h^{\gamma}_{\  \gamma}$ and all Lorentz 
indices are written as superscripts whereas
the derivatives are written as subscripts. All indices occurring twice are contracted by the Minkowski metric
$\eta^{\mu\nu}=\mathrm{diag}(1,-1,-1,-1)$. 
Since the perturbative expansion for the $S$-matrix is in powers of the coupling constant
$\kappa$, we are allowed to take for the first order 
cubic interaction between gravitons only the contribution
coming from ${\mathcal{L}}_{\sss HE}^{\sss (1)}$.

After quantization, Eq.~(\ref{eq:12}), the coupling~(\ref{eq:13}), completed by a suitable ghost-graviton
coupling term (see Sec.~\ref{sec:struttura}), can be used in perturbation theory to calculate quantum corrections
to classical general relativity.

Two serious problems arise in this procedure. 
The first one is the \textit{non-renormalizability of quantum gravity} due
to presence of two derivatives on the graviton fields in~(\ref{eq:13}) whose origin lies in the dimensionality
of the coupling constant ($[\kappa ]=\mathrm{mass}^{-1}$).
The second one is the non-polynomial character of $\mathcal{L}_{\sss HE}$, Eq.~(\ref{eq:11}), which reflects itself
into a \textit{proliferation of couplings}, \emph{i.e.} into an increasing polynomial degree in the interaction structure.

The first drawback, non-renormalizability, can be  approached  by means of the inductive causal 
construction of the $T_{n}$'s, which makes it possible to find finite and cutoff-free 
quantum corrections  for any process describable in the $S$-matrix framework, although the solution is
not quite clear with regard to physical predictability due to  the increasing number of finite
normalization terms in the distribution splitting~(\ref{eq:7}) in each order of perturbation theory.

With regard to the second issue, we could try to generalize the  result of~\cite{schorn1} and the more
recent result of~\cite{Schawell}, which suggest that
the concept of \textit{perturbative quantum operator gauge invariance}  (see Sec.~\ref{sec:struttura}) may be able to explain the
higher polynomial couplings:
gauge invariance to second order  will  automatically imply  the introduction of a  quartic graviton interaction 
exactly as prescribed by the expansion of the Hilbert--Einstein Lagrangian~\cite{schorn1}, see Sec.~\ref{sec:tree}.  
If we were able to repeat 
this step in each order of perturbation theory, we would  recover the full Einstein gravity in quantum form. At this
point the somewhat artificial decomposition of the metric tensor into a flat background and a dynamical variable 
would acquire a merely \textit{book-keeping} purpose beside the fact that we consider an asymptotically flat situation.

\section{Gauge Structure of Quantum Gravity}
\label{sec:struttura}
\setcounter{equation}{0}

\subsection{Gauge Charge}

The classical gauge transformations $h^{\alpha\beta} \to h^{\alpha\beta}
+u^{\alpha ,\beta}+u^{\beta ,\alpha} - \eta^{\alpha\beta} u^{\sigma}_{,\sigma}$, which corresponds 
to the linearized  general covariance of $g^{\alpha\beta}(x)$~\cite{wald},
can be implemented on a quantum level by means of the \textit{gauge charge} $Q$
\begin{gather}
h^{'\alpha\beta}(x)=e^{-i\lambda Q} h^{\alpha\beta}(x) e^{+i\lambda Q}\ ,\nonumber \\
Q:=\int\limits_{x^{0} =const}\!\! d^{3}x\, h^{\alpha\beta}(x)_{,\beta} {\stackrel{\longleftrightarrow}{ \partial_{0}^{x} }}
 u_{\alpha}(x)\, .
\label{eq:14}
\end{gather}
In order to get a nilpotent  gauge charge ($Q^2 =0$, in order to prove unitarity of the $S$-matrix and to construct the physical subspace
of the graviton Fock space), we have to quantize
the vector field $u^{\mu}(x)$, the \textit{ghost field} ($\Box u^{\nu}(x)=0$), with its partner $\tilde{u}^{\nu}(x)$, 
the \textit{anti-ghost field} (with $\Box \tilde{u}^{\nu}(x)=0$, too), as free fermionic vector fields through the anti-commutator
\begin{equation}
\big\{u^{\mu}(x),\tilde{u}^{\nu}(y)\big\}=i\, \eta^{\mu\nu}\, D_{0}(x-y) \, ,
\label{eq:g0}
\end{equation}
whereas all other anti-commutators vanish.

The gauge charge $Q$ defines an \textit{infinitesimal gauge variation} by
\begin{equation}
d_Q A:= Q\, A-(-1)^{n_{\sss{G}} (A)} A\, Q\, ,
\label{eq:14.2}
\end{equation}
where $n_{\sss{G}} (A)$ is the number of ghost fields minus the number of anti-ghost fields in the Wick monomial $A$. The operator
$d_Q$ obeys also the Leibniz rule
\begin{equation}
d_Q (AB)=(d_Q A)\, B +(-1)^{n_{\sss{G}} (A)} A\,  d_Q B \, ,
\end{equation}
for arbitrary operators $A$ and  $B$.

The infinitesimal operator gauge variations of the fundamental asymptotic free quantum fields are
\begin{gather}
d_Q h^{\alpha\beta}(x)  =\big[ Q, h^{\alpha\beta}(x)\big]=-i\, b^{\alpha\beta\rho\sigma} u^{\rho}(x)_{,\sigma}\, ,\nonumber\\
d_Q u^{\alpha}(x) = \big\{ Q, u^{\alpha}(x)\big\}=0\, , \nonumber \\
d_Q \tilde{u}^{\alpha}(x) = \big\{ Q, \tilde{u}^{\alpha}(x)\big\}=i\, h^{\alpha\beta}(x)_{,\beta}\, .
\label{eq:14.3}
\end{gather}

\subsection{Perturbative Operator Quantum Gauge Invariance}
\label{sec:gauge}

Formally, S-matrix gauge invariance means $\lim_{g\to 1} d_{Q}S(g)=0$. This follows from
\begin{equation} 
\lim_{g \to 1}\bigl( S'(g)-S(g)\bigr)=\lim_{g \to 1}\left(
-i\lambda \left[ Q,S(g)\right]+\hbox{higher commutators}\right)=0\,,
\end{equation}
which holds true, if the condition of \textit{perturbative gauge invariance to $n$-th order}
\begin{equation}
\begin{split}
d_{Q}T_{n}(x_{1},\ldots , x_{n}) =\left[ Q,T_{n}(x_{1},\ldots , x_{n})\right]
=\hbox{sum of divergences ,} 
\label{eq:16}
\end{split}
\end{equation}
is fulfilled  for all $ n \ge 1$.

\subsection{Gauge Invariance to First Order}
\label{sec:pert}

Already for $n=1$, Eq.~(\ref{eq:16}) is non-trivial,
because $d_{Q}T_{1}^{h, \sss HE}(x)\neq divergence $. This requires the introduction of a  anti-ghost--graviton--ghost
coupling~\cite{schorn1}
\begin{equation}
\begin{split}
T_{1}^{u,\sss KO}=i\kappa\big(&+:\tilde{u}^{\nu}(x)_{,\mu} h^{\mu\nu}(x)_{,\rho} u^{\rho}(x):
           -:\tilde{u}^{\nu}(x)_{,\mu}h^{\nu\rho}(x) u^{\mu}(x)_{,\rho}: \\
           &  - :\tilde{u}^{\nu}(x)_{,\mu}h^{\mu\rho}(x)u^{\nu}(x)_{,\rho}:
                    +:\tilde{u}^{\nu}(x)_{,\mu}h^{\mu\nu}(x)u^{\rho}(x)_{,\rho}:
           \big)\, ,
\label{eq:17}
\end{split}
\end{equation}
which was first derived by Kugo and Ojima in ~\cite{Kugo, Nishi}. 
Therefore,  we obtain
\begin{equation}
d_{Q}\left(T_{1}^{h,\sss HE}(x)+T_{1}^{u, \sss KO}(x)\right)=:
\partial_{\nu}^{x}T_{1/1}^{\nu}(x)=\hbox{sum of  divergences .}
\label{eq:18}
\end{equation}
One explicit  form of $T_{1/1}^{\nu}\sim\big\{:\!uhh\!: + :\!\tilde{u}uu\!:\big\}^{\nu}$, the so-called $Q$-vertex, 
was derived in~\cite{schorn1}.
The ghost couplings in causal quantum gravity are analyzed in great detail in~\cite{schorn2}.

The fermionic quantization of the ghost fields, usually called \textit{Faddeev-Popov ghosts}, is not only necessary for having 
a nilpotent  $Q$, but also for perturbative gauge invariance to be fulfilled.

In the path-integral framework, the ghost fields appear as a consequence of 
the quantization after gauge fixing~\cite{Faddeev}, but it was already noticed
by Feynman~\cite{Feynman} that without ghost fields a unitarity breakdown occurs in second
order at the loop level.

Although the condition $d_{Q}T_{1}(x)=divergence$ seems to be rather easy to fulfil, it has two important
consequences.
First of all, it rules out the possibility of a renormalizable theory of quantum gravity~\cite{Schawell}, because for a 
renormalizable 
interaction $T_{1}(x)$, \emph{i.e.} without the two derivatives acting on the 
fields\footnote{with only one derivative it is
impossible to form a Lorentz scalar interaction term}, perturbative gauge invariance to first order
entails  only the trivial solution $T_{1}(x)=0$.

The other interesting consequence pointed out  in~\cite{Schawell} is the following: if 
$T_{1}^{h+u}(x)$ is  the most general ansatz for the graviton coupling and  the most general ansatz for the ghost coupling
\begin{equation}
T_{1}^{h+u}(x)=\sum_{j} a_{j}\,:\big\{\!h h h\!\big\}_{j}:+\sum_{j} b_{j}\,:\big\{\!\tilde{u} h u \!\big\}_{j}:\,,
\end{equation}
(with two derivatives acting on the fields), then the requirement  $d_{Q}T_{1}^{h+u}(x)=divergence$
selects a small number of possible theories and the Hilbert--Einstein graviton coupling $T_{1}^{h,\sss HE}$,
with  the Kugo--Ojima ghost coupling $T_{1}^{u, \sss KO}$, lies among them.  Moreover, all the allowed 
couplings can be transformed in such a way that the most general coupling has now the form
\begin{equation}
T_{1}(x)=T_{1}^{h,\sss HE}(x)+ T_{1}^{u,\sss KO}(x) + \text{divergence couplings}+d_{Q}(\tilde{u}hh +\tilde{u}\tilde{u}u)\, .
\end{equation}
The last term represents the so-called \textit{coboundary terms} which, together with divergence terms, seem to play
no physical r\^ ole.

The definition of the $Q$-vertex from Eq.~(\ref{eq:18}) allows us to give  a precise prescription on how the right side of
Eq.~(\ref{eq:16}) has to be inductively constructed.
We define the concept of \textit{perturbative quantum operator gauge invariance} by the equation
\begin{equation}
d_{Q}T_{n}(x_{1},\ldots ,x_{n})=\sum_{l=1}^{n}\frac{\partial}{\partial x^{\nu}_{l}}\, 
T_{n/l}^{\nu}(x_{1},\ldots , x_{l},\ldots ,x_{n}) \, .
\label{eq:37.1}
\end{equation}
Here, $T_{n/l}^{\nu}$ is the time-ordered renormalized product, obtained according to the inductive causal scheme, with a
$Q$-vertex at $x_{l}$, while all other $n-1$ vertices are ordinary $T_{1}$-vertices.

Analysis of the condition~(\ref{eq:37.1}) shows that perturbative gauge invariance can be spoiled by local 
terms, \emph{i.e.}
terms proportional to $:\!\mathcal{O}(x_{1},\ldots ,x_{n})\! : \delta^{\sss (4n-4)}(x_{1}-x_{n},\ldots ,x_{n-1}-x_{n})$, which
may appear as a consequence of distribution splitting on both  sides  of Eq.~(\ref{eq:37.1}).

If it is possible to absorb these local terms by suitable local normalization terms  $N_{n}$ of
$T_{n}$ and $N_{n/l}^{\nu}$ of $T_{n/l}^{\nu}$ in such a way that the equation
\begin{equation}
d_{Q}\Big(T_{n}+N_{n}\Big)(x_{1},\ldots ,x_{n})=\sum_{l=1}^{n}\frac{\partial}{\partial x^{\nu}_{l}}\, 
\Big( T_{n/l}^{\nu}+N_{n/l}^{\nu} \Big) (x_{1},\ldots , x_{l},\ldots ,x_{n}) 
\label{eq:38.1}
\end{equation}
holds true, then we call the theory gauge invariant to $n$-th order.

\section{Quantum Gravity in Second  Order}
\label{sec:causal}
\setcounter{equation}{0}

Before undertaking the examination of the various contributions in second order perturbation theory
(tree, self-energy and vacuum graphs), we give the formula for the singular order of arbitrary $n$-point distributions
in perturbative quantum gravity.

We consider in the $n$-th order of perturbation theory an arbitrary $n$-point distribution 
$T^{\sss G}_{n}(x_{1},\ldots,x_{n})$, 
appearing in Eq.~(\ref{eq:1}), as a sum of normally ordered products of free field operators multiplied by numerical 
distributions
\begin{equation}
T^{\sss G}_{n}(x_{1},\ldots ,x_{n})
= :\!\prod_{j=1}^{n_h}h(x_{k_{j}})\,\prod_{i=1}^{n_{u}}u(x_{m_{i}})\,\prod_{l=1}^{n_{\tilde{u}}}
\tilde{u}(x_{n_{l}})\!:\,t^{\sss G}_{n}(x_{1},\ldots ,x_{n})\,.
\label{eq:b38}
\end{equation}
This $T_{n}^{\sss G}$ corresponds to a graph $G$ with $n_{h}$ external graviton lines , $n_{u}$ external ghost lines  
and $n_{\tilde{u}}$ external anti-ghost lines. 

The singular order of $G$ then reads
\begin{equation}
\omega(G)\le 4-n_{h}-n_{u}-n_{\tilde{u}}-d+n\,.
\label{eq:b39}
\end{equation}
Here $d$ is the number of derivatives on the external field operators in~(\ref{eq:b38}). The $\le$ means that in 
certain cases the singular order is lowered by peculiar conditions, \emph{e.g.} by the equations of motions of  free 
fields.

In the usual QFT formulation, Eq.~(\ref{eq:b39}) implies the \textit{non-renormalizability} of QG,
because  $\omega(G)$ increases without bound for higher  orders in the perturbative expansion.
This means that there is a \textit{proliferation} of divergences and of counterterms (one still has to hope that
the needed counterterms can be fitted into the original Lagrangian) to remove  them.

The hope that QG was UV-finite to all orders failed  after the two-loop calculation in~\cite{Goroff1,Goroff2,vandeVen1},
although the one-loop order is UV-finite ~\cite{Hooft,deser2}.

The situation is different in causal perturbation theory: we are facing in this case  a \textit{non-normalizable} 
theory.  The theory has  a weaker predictive power but it is still well-defined in the sense of UV finiteness.

The ambiguity in the normalization reflects itself into an  increasing number of  free, undetermined but  finite
constants in  Eq.~(\ref{eq:7}).
The problem is then  to find enough physical conditions or  requirements to fix  this increasing  freedom 
and to investigate the effects of these local interactions for physical quantities.

\subsection{Graviton Self-Energy}
\label{sec:loop}

We investigate the graviton self-energy contribution (graviton and ghost
loops) in second order. As in Sec.~\ref{sec:smatrix}, the inductive construction of $T_{2}(x_{1},x_{2})$ can
be accomplished in two steps: in the first place  we construct the  causal
distribution $D_{2}$  from Eq.~(\ref{eq:13}) and (\ref{eq:17}) by applying
Wick expansion with the contractions given by Eq.~(\ref{eq:12})
\begin{equation}
\begin{split}
D_{2}^{\sss SE}(x_{1},x_{2})&=\Big[ T_{1}^{h +u}(x_{1}),
 T_{1}^{h + u}(x_{2})\Big]\bigg|_{\sss SE}\\
&=:\!h^{\alpha\beta}(x_{1})h^{\mu\nu}(x_{2})\!:
d^{\sss SE}_{2}(x_{1}-x_{2})_{\alpha\beta\mu\nu}\,.
\label{eq:19}
\end{split}
\end{equation}
Because of translation invariance the C-number distribution $d^{\sss SE}_{2}$ depends
only on the relative coordinate $x=x_{1}-x_{2}$. In momentum space we obtain
for the self-energy tensor
\begin{equation}
\hat{d}_{2}^{\sss SE}(p)_{\alpha\beta\mu\nu}=
\hat{P}(p)_{\alpha\beta\mu\nu}^{\sss (4)}\, 
\Theta(p^2)\, \textrm{sgn}(p^{0})\,,
\label{eq:20}
\end{equation}
where $\hat{P}(p)_{\alpha\beta\mu\nu}^{\sss (4)}$ is a covariant polynomial of
degree 4
\begin{equation}
\begin{split}
\hat{P}(p)_{\sss (4)}^{\alpha\beta\mu\nu}=\frac{\kappa ^2 \pi}{960 (2\pi)^4}
\bigg[
&-656\:  p^{\alpha}p^{\beta}p^{\mu}p^{\nu}
 -208\:  p^2\big( p^{\alpha}p^{\beta}\eta^{\mu\nu}+p^{\mu}p^{\nu}\eta^{\alpha\beta}\big)\\
&+162\:  p^2 \big(p^{\alpha}p^{\mu}\eta^{\beta\nu}+p^{\alpha}p^{\nu}\eta^{\beta\mu}+p^{\beta}p^{\mu}\eta^{\alpha\nu}+
                   p^{\beta}p^{\nu}\eta^{\alpha\mu}\big) \\
&-162\:  p^4 \big(\eta^{\alpha\mu}\eta^{\beta\nu}+\eta^{\alpha\nu}\eta^{\beta\mu}\big)
 +118\:  p^4 \eta^{\alpha\beta}\eta^{\mu\nu}\bigg]\, .
\label{eq:21}
\end{split}
\end{equation}
Then, in order to obtain $T^{\sss SE}_{2}(x_{1},x_{2})$, we split
$d^{\sss SE}_{2}(x)$  according to the singular order $\omega(d^{\sss SE}_{2})=4$, obtained 
from Eq.~(\ref{eq:b39}) or from direct inspection of Eq.~(\ref{eq:21}).
Thus, admitting free normalization polynomial terms $\hat{N}(p)_{(2a)}^{\alpha\beta\mu\nu}$, which
correspond to local interaction terms in configuration space, we obtain 
\begin{gather}
T^{\sss SE}_{2}(x_{1},x_{2})=:\!h^{\alpha\beta}(x_{1})h^{\mu\nu}(x_{2})\!: \,i\, 
\Pi (x_{1}-x_{2})^{tot}_{\alpha\beta\mu\nu}\, ,\nonumber \\
\hat{\Pi}(p)_{tot}^{\alpha\beta\mu\nu}=
\frac{\hat{P}(p)_{\sss (4)}^{\alpha\beta\mu\nu}}{2\pi}\, \log\left(\frac{-(p^2+i0)}{M_{0}^2}\right)+
\sum_{a=0}^{2}\hat{N}(p)_{(2a)}^{\alpha\beta\mu\nu} \, .
\label{eq:22}
\end{gather}
The scalar distribution
\begin{equation}
\hat{t}(p)=\frac{i}{2\pi}\log\left(\frac{-(p^2+i0)}{M_{0}^2}\right)\, ,
\end{equation}
is calculated  from the massless causal scalar 
distribution $\hat{d}(p)=\hat{r}'(p)-\hat{a}'(p)=\Theta(p^2)\, 
\textrm{sgn}(p^{0})$ in Eq.~(\ref{eq:20})  by splitting it into $\hat{d}(p)=\hat{r}(p)-\hat{a}(p)$ 
and subtracting $\hat{r}'(p)$ from $\hat{r}(p)$, see~\cite{ym2,gri4}. The mass scale  $M_{0}$ represents a normalization constant
and  not a cutoff.

To get a condition for the undetermined normalization terms, we consider the sum of the proper 
self-energy diagrams with an increasing number of self-energy insertions.
By requiring that the mass of the graviton (which is zero) and the coupling constant $\kappa$ remain 
unchanged under these radiative corrections, we find that all normalization terms must vanish, 
except for the term of degree 4 which
can be absorbed in the new parameter $M_{0}^2$, see~\cite{gri4} for details.

We emphasize the fact that, in
virtue of the causal splitting prescription, all expressions are UV-finite and
Eq.~(\ref{eq:22}) agrees exactly with the finite part obtained using standard  
regularization schemes~\cite{Capper,Zaidi}. As a consequence it is not necessary to add counterterms~\cite{Hooft}
to the original Lagrangian in order to  renormalize the theory. 

The graviton self-energy 
satisfies the Slavnov--Ward identity for the $2$-point connected Green function~\cite{Capper,Zaidi}
\begin{equation}
p^{\alpha}p^{\mu}\big\{ b_{\alpha\beta\gamma\delta}
\big[ \hat{\Pi}^{{\sss{SE}}, grav.}(p)^{\gamma\delta\rho\sigma}+
\hat{\Pi}^{{\sss{SE}}, ghost}(p)^{\gamma\delta\rho\sigma}\big]
b_{\rho\sigma\mu\nu}\big\} =0\, ,
\label{eq:23}
\end{equation}
only if ghost and graviton loops are taken into account,
as well as perturbative gauge invariance, Eq.~(\ref{eq:16}): $d_{Q}T_{2}^{\sss SE}
(x_{1},x_{2})=divergence$.

The result of Eq.~(\ref{eq:22})  can also  be used to find the long range, low energy quantum corrections
to the Newtonian  potential 
between two bodies of mass  $m_{1}$ and $m_{2}$ at a distance $r$ in the non-relativistic static limit
(see Sec.~\ref{sec:matter} for the coupling between matter and gravity). 

In the spirit of~\cite{Dono1,Dono2,Hamber}, but without resorting to any effective field theore\-tical calculation,
we compute a matter--matter scattering diagram with exchange of one graviton. The corresponding scattering amplitude
leads to the Newtonian potential $V(r)=-G\,m_{1} m_{2}\,r^{-1}$.

 Considering also the radiative corrections coming
from the graviton self-energy, we obtain  quantum  corrections to the Newtonian potential. More precisely,
 we find that the logarithm depending term in Eq.~(\ref{eq:22}) gives the  $r^{-3}$-correction
\begin{equation}
V(r)=\frac{-G\,m_{1}\,m_{2}}{r}\,\bigg( 1 +\frac{206}{30}\, \frac{ G\, \hbar}{c^3\,\pi\, r^2}\bigg)\,.
\label{eq:24}
\end{equation}
In Sec.~\ref{sec:matter}, 
when we consider also scalar massless  matter, we will find supplementary corrections coming 
from these massless particle loops.

The central  piece in the calculation is the distributional Fourier transform of $\log\left(\bol{p}^{2}/M_{0}^{2}\right)$
which yields $(-2\pi r^3)^{-1}$ and the $M_{0}$-dependence disappears from the non-local part of the final result,
being proportional to $\delta^{\sss (3)}(\bol{x})$ with $r=|\bol{x}|$. 

The relevant length scale appearing in Eq.~(\ref{eq:24}) 
is the Planck  length $\ell_{pl.}=\sqrt{G\hbar/c^{3}}$. Therefore, appreciable quantum corrections manifest themselves  only for
$r\sim \ell_{pl.}$.

Our result agrees with the corresponding one in~\cite{Hamber}, although this represents only a partial correction to the 
Newtonian potential, because we have taken into account only the graviton self-energy contribution and not the complete 
set of diagrams of order  $\kappa^4$  contributing to these corrections, as, for example, the vertex correction or the double 
scattering. Therefore we cannot make any statement on the absolute sign of the correction in Eq.~(\ref{eq:24}).

\subsection{Tree Graphs and Gauge Invariance}
\label{sec:tree}

For the tree graphs we quote briefly the result of Schorn~\cite{schorn1,schorn2}: \textit{perturbative gauge 
invariance to second  order  \textsl{generates} the 4-graviton coupling}.

The condition of perturbative gauge invariance to second order
\begin{equation}
d_{Q}T_{2}(x,y)=\d_{\sigma}^{x}\,T_{2/1}^{\sigma}(x,y)+\d_{\sigma}^{y}\,T_{2/2}^{\sigma}(x,y)
\label{eq:a1}
\end{equation}
restricted to the operator structure $:\!uhhh\!:$ and $:\!\tilde{u}uuh\!:$ can be spoiled by terms with local support, 
namely proportional to $\delta^{\sss (4)}(x-y)$: local normalization terms $N_{2}$, $N^{\sigma}_{2/1}$ and  $N^{\sigma}_{2/2}$ 
of $T_{2}$, $T^{\sigma}_{2/1}$ and $T^{\sigma}_{2/2}$ respectively, and local 
anomalies\footnote{Anomalies are  terms arising from $\d_{\sigma}^{x}\,T_{2/1}^{\sigma}(x,y)+\d_{\sigma}^{y}\,T_{2/2}^{\sigma}(x,y)$
because of   the following mechanism:
\begin{equation*}
\begin{split}
\d_{\sigma}^{x} \big(:\!\mathcal{O}(x,y)\!:\,\d_{x}^{\sigma} D_{0}^{\sss F}(x-y)\big)=&\ldots+
:\!\mathcal{O}(x,y)\!:\Box D_{0}^{\sss F} (x-y)\\
=& \ldots+\underbrace{:\!\mathcal{O}(x,y)\!: \delta^{\sss (4)}(x-y)}_{\text{local anomaly}}
\end{split}
\end{equation*}}.
Therefore, we have to investigate the equation
\begin{equation}
d_{Q}N_{2}(x,y)=\mathrm{an}\big(\d_{\nu}^{x}T_{2/1}^{\nu}(x,y)+
\d_{\nu}^{y}T_{2/2}^{\nu}(x,y)\big)
+\d_{\nu}^{x}N_{2/1}^{\nu}(x,y)+
\d_{\nu}^{y}N_{2/2}^{\nu}(x,y)\,,
\label{eq:xx}
\end{equation}
which relates the local terms appearing in Eq.~(\ref{eq:a1}). 

We take advantage of  the freedom in the normalization of  tree diagrams with singular order $\omega\ge 0$
(they appear because of the two derivatives present in the coupling) by choosing local normalization terms $N_{2}(x,y)$ of the form
\begin{equation}
\begin{split}
N_{2}(x,y)=i\,\kappa^2 \,\Big\{
&+  :\!h^{\rho\sigma}h^{\alpha\beta}h^{\alpha\beta}_{,\rho}h_{,\sigma}\!:
 -2 :\!h^{\rho\sigma}h^{\alpha\beta}h^{\alpha\mu}_{,\rho}h^{\beta\mu}_{,\sigma}\!:
 +2 :\!h^{\rho\sigma}h^{\sigma\alpha}h^{\alpha\beta}_{,\mu}h^{\beta\rho}_{,\mu}\!:+\\
&-2 :\!h^{\rho\sigma}h^{\rho\mu}h^{\sigma\alpha}_{,\beta}h^{\mu\beta}_{,\alpha}\!:
 -\frac{1}{2} :\!h^{\rho\sigma}h^{\rho\sigma}_{,\mu}h^{\alpha\beta}h^{\alpha\beta}_{,\mu}\!:
 -  :\!h^{\rho\sigma}h^{\rho\alpha}h^{\alpha\sigma}_{,\beta}h_{,\beta}\!:+\\
&+:\!h^{\rho\sigma}h^{\sigma\alpha}_{,\mu}h^{\alpha\beta}h^{\beta\rho}_{,\mu}\!:
\Big\}\,\delta^{\sss(4)}(x-y)\,,
\end{split}
\end{equation}
for  Eq.~(\ref{eq:xx}) to be fulfilled, see~\cite{schorn1} for details.

Taking the factor $1/2$ for the second order of the $S$-matrix 
expansion   into account, these  quartic interactions (quadratic 
in $\kappa$)  agree exactly with the terms of order $\kappa^2$ 
in the expansion of the Hilbert--Einstein  Lagrangian density 
$\mathcal{L}_{\sss EH}$ given by Eq.~(\ref{eq:11}).
 This mechanism of  generation of the higher orders works 
in a purely quantum framework.

Such a property was already observed in Yang--Mills theories~\cite{ym1}: starting with an interaction 
between three gauge fields, perturbative gauge invariance generates automatically the 4-gauge fields coupling.

Since QG is constructed starting from  a non-polynomial Lagrangian, it is not clear if this scheme would also 
work in higher orders and the question whether  perturbative gauge invariance to $n$-th order 
requires the introduction of local terms which turn out to agree with the $(n+1)$-th  term in the expansion of 
the Hilbert--Einstein  Lagrangian remains unanswered.

\subsection{Vacuum Graphs in Second Order}
\label{subsec:vac}

We discuss also the vacuum graphs in second order.
In the causal perturbation theory they cannot be \textit{divided away}
as in the Gell--Mann and Low  series for connected Green functions, but this is not a  problem
because they are finite as a consequence of the causal scheme. The corresponding distribution $T_{2}^{\sss VG}$
has been obtained by computing three contractions in Eq.~(\ref{eq:bb}). It  has singular
order $\omega =6$ and reads
\begin{equation}
\hat{T}_{2}^{\sss VG}(p)=\frac{i\, \kappa ^2\, \pi ^2}{(2\pi)^8}\,\frac{-37}{512}\,  p^6\,
\log\left(\frac{-(p^2+i0)}{M_{0}^{2}} \right)
+\sum_{i=0}^{3} c_{i}(p^2)^{i}\, .
\label{eq:a3}
\end{equation} 
It is possible to prove the free vacuum stability in QG as described in~\cite{Scha2}:
$\lim_{g\uparrow 1}\bigl(\Omega ,S(g)\Omega\bigr)=1$, where $\Omega$
is the Fock vacuum of free asymptotic fields. 
Perturbatively this means
$\lim_{g\uparrow 1}\bigl(\Omega ,S_{n}(g)\Omega\bigr)=0\,\, , \forall n\ge 1$.

We perform the adiabatic limit in scaling form: $ g(x)=g_{0}(\epsilon x)$ where $\epsilon \to 0$ and 
$g_{0}\in \mathcal{S}(\mathbb{R}^{4})$ with $g_{0}(0)=1$.
For $n=2$ we get
\begin{equation*}
\lim_{g\to  1}\big(\Omega ,S_{2}(g)\Omega\big)=\frac{(2\pi)^{2}}{2}\lim_{\epsilon\to 0}
\frac{1}{\epsilon ^{4}}\int\!\!d^{4}p\Big[
\underbrace{ \hat{T}_{2}^{\sss VG}(\epsilon p)}_{\sim (\epsilon p)^{6}}+
\hat{N}_{2}^{\sss VG}(\epsilon p)\Big]\ \hat{g}_{0}(p)\,  \hat{g}_{0}(-p)=0\, ,
\label{eq:a4}
\end{equation*}
as a consequence of the
\textit{bad} UV behaviour of QG ($\hat{T}_{2}^{\sss VG}(p)\sim p^{6}$). 

At the same time, free vacuum 
stability forces the free normalization constants $c_{i}$ to vanish. This allows the graviton to show up as an
asymptotic particle carrying the long range gravitational interaction.

\section{The Physical Subspace ${\mathcal{F}_{phys}}$}
\label{sec:physical}
\setcounter{equation}{0}

An interesting feature of this approach to the gauge structure of QG is the   construction of  the physical Hilbert--Fock 
space for the  asymptotic free graviton field. 

In order to decouple the ghosts and the unphysical
degrees of freedom of the graviton from the truly physical degrees of freedom in the theory, we could apply the
Gupta-Bleuler~\cite{gupta1} formalism with indefinite metric, but we prefer  to realize
the free field  representations on a Fock space  with positive definite metric~\cite{Aste2}. 
Lorentz covariance requires then the introduction of a  Krein
structure~\cite{Kra,ym4} on the Fock space  and we can characterize  the physical subspace 
${\mathcal{F}}_{phys}$  by the following definition
\begin{equation}
{\mathcal{F}}_{phys}:=\ker\left\{Q,Q^{\dagger}\right\}\,  .
\label{eq:b4}
\end{equation}
In order to verify the consistency of this formula, we need an explicit representation of the free fields appearing in the theory. 

Since a symmetric tensor field with arbitrary trace transforms under the proper Lorentz group $\mathcal{L}_{+}^{\uparrow}$ 
according to the tensor product of two spinor representations ${\mathcal D}^{(1/2,1/2)}$, we decompose  
$ h^{\alpha \beta}(x)$ according 
to the irreducible reduction of the representations
\begin{equation}
{\mathcal D}^{(1/2,1/2)} \otimes {\mathcal D}^{(1/2,1/2)} \vert _{sym}
={\mathcal D}^{(1,1)}\oplus  {\mathcal D}^{(0,0)}\, ,
\label{eq:b5}
\end{equation}
into
\begin{equation}
h^{\alpha \beta}(x)=H^{\alpha\beta}(x) +\frac{1}{4} \eta^{\alpha\beta}\Phi(x)\,,
\label{eq:b6}
\end{equation}
where $ H^{\alpha\beta}(x)$ represents a traceless symmetric tensor field defined as 
$H^{\alpha\beta}(x):=h^{\alpha\beta}(x)- \eta^{\alpha\beta}h(x)/4$ with $H^{\gamma}_{\ \gamma}=0$
(9 degrees of freedom) and  $ \Phi(x)$ a scalar field with $h^{\gamma}_{\ \gamma}=\Phi$.
From Eq.~(\ref{eq:12}) we obtain the following commutation relations
\begin{gather}
\big[ \Phi(x),\Phi(y)\big]=4\, i\, D_{0}(x-y)\, ,\quad  \big[ H^{\alpha\beta}(x),\Phi(y) \big]=0\,,   \nonumber\\
\big[ H^{\alpha\beta}(x),H^{\mu\nu}(y)\big] = - \frac{i}{2} \big(   \eta^{\alpha\mu} \eta^{\beta\nu} +
\eta^{\alpha\nu} \eta^{\beta\mu}  - \frac{1}{2} \eta^{\alpha\beta} \eta^{\mu\nu} \big)
   \, D_{0}(x-y)\nonumber\\
=-i\, t^{\alpha\beta\mu\nu}\, D_{0}(x-y)\, .
\end{gather}

For the quantization of $H^{\alpha\beta}(x)$ and $\Phi(x)$ we choose the following free field representations
\begin{equation}
\begin{split}
H^{\alpha\beta}(x)&=(2\pi)^{-3/2}\int \frac{d^{3}k}{\sqrt{2\omega}}\Big( A^{\alpha\beta}(\mathbf{k})e^{-ikx}+
A^{\alpha\beta}(\mathbf{k})^
{\mathrm K} e^{+ikx}\Big)\, ,\\
\Phi(x)&=(2\pi)^{-3/2}\int \frac{d^{3}k}{\sqrt{2\omega}}\Big( a(\mathbf{k})e^{-ikx}+
a(\mathbf{k})^{\mathrm K} e^{+ikx}\Big)\, ,
\end{split}
\end{equation}
where the Krein conjugation  $\mathrm{K}$  is defined by the Krein operators $\eta_{\sss H}$ and
$\eta_{\sss \Phi}$
\begin{gather}
(A^{\alpha\beta})^{\mathrm{K}}=\eta_{\sss H}{ A^{\alpha\beta}}^{\dagger} \eta_{\sss H}\ , 
\quad 
\eta_{\sss H}=\bigotimes_{i=1}^{3} (-1)^{\mathbf{N}_{0i}}\ , \quad \mathbf{N}_{0i}=2 \int
 d^{3}k\, A^{0i}(\mathbf{k})^{\dagger} A^{0i}(\mathbf{k})\, ; \nonumber\\
(a)^{\mathrm{K}}=\eta_{\sss \Phi} a^{\dagger} \eta_{\sss  \Phi} \ , \quad 
\eta_{\sss \Phi}=(-1)^{\mathbf{N}_{\Phi}}\ , \quad \mathbf{N}_{\Phi}=
\frac{1}{4}\int d^{3}k\, a(\mathbf{k})^{\dagger}a(\mathbf{k})\, .
\end{gather}
The fields are then $\mathrm{K}$-selfconjugate  and not $\dagger$-selfadjoint, but the field components which spoil
the selfadjointness turn out to be the unphysical ones and therefore  these will be absent in the physical 
subspace ${\mathcal{F}}_{phys}$, so that on  ${\mathcal{F}}_{phys}$ one has ${H^{\alpha\beta}}^{\mathrm{K}}
={H^{\alpha\beta}}^{\dagger}$.
The absorption and creation  operators $A^{\alpha\beta}=A^{\beta\alpha}\ ,{A^{\alpha\beta}}^{\dagger}$  
and $a, a^{\dagger}$ satisfy the canonical
commutation relations
\begin{gather}
\big[ A^{\alpha\beta}(\mathbf{k}),A^{\mu\nu}(\mathbf{p})^{\dagger}\big]=
    \frac{1}{2}\big(\delta^{\alpha\mu}\delta^{\beta\nu}+\delta^{\alpha\nu}\delta^{\beta\mu}
-\frac{1}{2}\eta^{\alpha\beta}\eta^{\mu\nu}\big)\  \delta^{\sss (3)}
(\mathbf{k}-\mathbf{p})\nonumber\\ =\tilde{t}^{\alpha\beta\mu\nu}\ \delta^{\sss (3)}
(\mathbf{k}-\mathbf{p})\, ,          \nonumber     \\
\big[a(\mathbf{k}),a^{\dagger}(\mathbf{p})\big]=4\,\delta^{\sss (3)}
(\mathbf{k}-\mathbf{p})\, .
\label{eq:z1}
\end{gather}
The $\tilde{t}^{\alpha\beta\mu\nu}$-tensor has the following values
\vskip 5mm
\begin{center}
\begin{tabular}{|l|l|l|l|l|l|l|l|} 
\hline
$\tilde{t}^{\alpha\beta;\mu\nu}$ & $0 0; 0 0$ & $0 0; i i$ & $0 i; 0 i$ & $i i;
 i i$ & $i i; j j$ & $i j; i j$ & \text{otherwise}  \\   
\hline
value &  3/4  & 1/4  & 1/2  &    3/4  &  -1/4   &   1/2 & \:\: \:\  0 \\
\hline
\end{tabular}
\end{center}
\vskip 5mm
\noindent
with $i, j =1, 2, 3;\,  i \ne j$.
From this table we see that the $\tilde{t}^{\alpha\beta\mu\nu}$-tensor is neither diagonal nor positive definite,
although it is positive for the diagonal terms. In order to remedy these  defects, we define new absorption
operators
\begin{equation}
\begin{split}
A^{00} &= \frac{1}{2}(+ {a}^{11} +{a}^{22}   +{a}^{33} )\, , \quad
A^{11} = \frac{1}{2}(- {a}^{11} +{a}^{22}   +{a}^{33} )\, , \\
A^{22} &= \frac{1}{2}(+ {a}^{11} -{a}^{22}   +{a}^{33} )\, , \quad
A^{33} = \frac{1}{2}(+ {a}^{11} +{a}^{22}   -{a}^{33} )\, ;
\end{split}
\label{eq:a6}
\end{equation}
and analogously for the creation operators. Then we  obtain the commutation
relations
\begin{equation}
\big[ {a}^{ii}(\mathbf{k}),{a}^{jj}(\mathbf{p})^{\dagger}\big]=\delta^{ij}\, 
\delta^{\sss (3)}(\mathbf{k}-\mathbf{p})\, .
\end{equation}
Note that the operators ${a}^{00}$ and ${{a}^{00}}^{\dagger}$ do not appear here because this operator pair
is superfluous due to the trace condition $H^{\gamma}_{\ \gamma}=0$.

Now we want to specify the physical subspace with the help of the gauge charge $Q$,  which now reads
\begin{equation}
Q=\int\limits_{x^{0} = t}\!\! d^{3}x\, \Big( H^{\alpha\beta}(x)_{,\beta}+\frac{1}{4}\Phi(x)^{,\alpha}\Big)
 {\stackrel{\longleftrightarrow}{ \partial_{0}^{x} }}
 u^{\gamma}(x)\eta_{\alpha\gamma}   \, .
\end{equation}
For this purpose, we need the  free field representations of the ghost fields.
We follow in our discussion the analysis of the scalar ghost fields for Yang--Mills theories carried out in~\cite{Kra}.
Here we are dealing with vector ghost fields and 
we choose the following  free field representations
\begin{equation}
\begin{split}
u^{\nu}(x)&=(2\pi)^{-3/2}\int\!\frac{d^{3}p}{\sqrt{2\omega}}\Big( +b^{\nu}(\mathbf{p})e^{-ipx}
-\eta^{\nu\nu}{c^{\nu}(\mathbf{p})}^{\dagger} e^{ipx} \Big)\, ,\\
\tilde{u}^{\nu}(x)&=(2\pi)^{-3/2}\int\!\frac{d^{3}p}{\sqrt{2\omega}}\Big( -c^{\nu}(\mathbf{p})e^{-ipx}
-\eta^{\nu\nu}{b^{\nu}(\mathbf{p})}^{\dagger} e^{ipx} \Big)\, ;
\label{eq:g1}
\end{split}
\end{equation}
which satisfy the covariant commutation rule Eq.~(\ref{eq:g0}), whereas the absorption and creation operators
satisfy the commutation relations
\begin{gather}
\big\{ c^{\mu}(\mathbf{p}),c^{\nu}(\mathbf{k})^{\dagger}\big\} = \delta^{\mu\nu}\,  \delta^{\sss (3)}
(\mathbf{p} - \mathbf{k})\, ,\nonumber\\
\big\{ b^{\mu}(\mathbf{p}),b^{\nu}(\mathbf{k})^{\dagger}\big\} = \delta^{\mu\nu}\,  \delta^{\sss (3)}
(\mathbf{p} - \mathbf{k})\, .
\end{gather}
Extending the $\dagger$-conjugation  to the $\mathrm{K}$-conjugation we obtain the most symmetric form
\begin{equation}
\begin{split}
u^{\nu}(x)&=(2\pi)^{-3/2}\int\!\frac{d^{3}p}{\sqrt{2\omega}}\Big( +b^{\nu}(\mathbf{p})e^{-ipx}+
 {b^{\nu}(\mathbf{p})}^{\mathrm{K}}e^{ipx} \Big)\,  ,\\
\tilde{u}^{\nu}(x)&=(2\pi)^{-3/2}\int\!\frac{d^{3}p}{\sqrt{2\omega}}\Big( -c^{\nu}(\mathbf{p})e^{-ipx}
+{c^{\nu}(\mathbf{p})}^{\mathrm{K}} e^{ipx} \Big)\,  .
\label{eq:g2}
\end{split}
\end{equation}
This  implies $\big( u^{\nu}\big)^{\mathrm{K}}= u^{\nu}$ and $ \big( \tilde{u}^{\nu}\big)^{\mathrm{K}}=- \tilde{u}^{\nu}$,
or equivalently: $\left( b^{i}\right)^{\mathrm{K}} ={c^{i}}^{\dagger}$, $\left( c^{i}\right)^{\mathrm{K}} =
{b^{i}}^{\dagger}$, $\left( b^{0}\right)^{\mathrm{K}} =-{c^{0}}^{\dagger}$ and 
$\left( c^{0}\right)^{\mathrm{K}} =-{b^{0}}^{\dagger}$.

The construction of the Krein operator $\eta_{\sss G}$ on the ghost Fock space
which generates the transformations  $(O)^{\mathrm{K}}=\eta_{\sss G}O^{\dagger}\eta_{\sss G}$
from the $\dagger$-conjugation to the K-conjugation requires more work, see~\cite{gri4},
and  reads
\begin{equation}
\eta_{\sss G}=\exp\Big(i \frac{\pi}{2}\left(N_{g}-\Gamma_{g} \right) \Big)
\end{equation}
where 
\begin{gather}
N_{g}=N_{g}^{(0)}-\sum_{i=1}^{3}N_{g}^{(i)}\qquad \mathrm{with}\qquad  N_{g}^{(\mu)}=
          \int \!\!d^{3}p\, \big({b^{\mu}}^{\dagger}b^{\mu}+{c^{\mu}}^{\dagger}c^{\mu}\big)\,,\qquad \mathrm{and}\nonumber \\
\Gamma_{g}=-\sum_{\mu=0}^{3}\Gamma_{g}^{(\mu)} \qquad \mathrm{with}\qquad \Gamma_{g}^{(\mu)}=
     \int \!\!d^{3}p\, \big({b^{\mu}}^{\dagger}c^{\mu}+{c^{\mu}}^{\dagger}b^{\mu}\big)\, .
\end{gather}
$N_{g}^{(\mu)}$ represents the the $\mu$-ghost number operator whereas  $\Gamma_{g}^{(\mu)}$ represents the  $\mu$-ghost 
transfer  operator.

Calculating $ \left\{ Q,Q^{\dagger}\right\}$ in momentum space, we obtain
\begin{equation}
\begin{split}
\left\{Q,Q^{\dagger} \right\}=\int \! d^{3}k\, \omega^{2}(\mathbf{k}) \bigg[&
+2\sum_{\mu=0}^{3} \Big( {A^{0\mu}}(\bol{k})^{\dagger} A^{0\mu}(\bol{k}) +
{A^{\mu}_{\parallel}}(\bol{k})^{\dagger} A^{\mu}_{\parallel}(\bol{k}) \Big)+\\ 
&+\frac{1}{4}\, {a^{\dagger}(\bol{k})a(\bol{k})}
+\sum_{\alpha=0}^{3} \Big({ c^{\alpha}}(\bol{k})^{\dagger} c^{\alpha}(\bol{k}) +{ b^{\alpha}}(\bol{k})^{\dagger} b^{\alpha}(\bol{k})\Big)
+\\
&+\frac{1}{2}
\big(\sum_{\alpha=0}^{3} \frac{k_{\alpha}}{k_{0}} {c^{\alpha}}^{\dagger}(\bol{k})\big) \big(\sum_{\beta=0}^{3} \frac{k_{\beta}}{k_{0}} 
c^{\beta}(\bol{k})\big)+\\
&+\frac{1}{2} \big(\sum_{\alpha=0}^{3} \frac{k^{\alpha}}{k_{0}} {b^{\alpha}}^{\dagger}(\bol{k})\big)
 \big(\sum_{\beta=0}^{3} \frac{k^{\beta}}{k_{0}} b^{\beta}(\bol{k})\big)\bigg]\,,
\label{eq:a7}
\end{split}
\end{equation}
where
$A^{\mu}_{\parallel}$ represent the absorption operator for the $\mu$-longitudinal mode
\begin{equation}
A^{\mu}_{\parallel}(\bol{k}):=\frac{k_{i}}{\omega (\bol{k})}\,A^{\mu i}(\bol{k})\,.
\end{equation}
Apparently there is an over-counting in the graviton sector: we have  four $0\mu$- and four $\mu$-longitudinal 
modes number operators, as well as the 
scalar component $a$ number operator, but $A^{0}_{\parallel}$ is not independent, being a linear combination of the 
$A^{0i}$-operators and we have not taken into account that $\eta_{\alpha\beta}A^{\alpha\beta}=0$.

For this purpose let us choose a reference frame in which $k^{\mu}=(\omega,0,0,\omega)$ is parallel to the 
third axis, because obviously the unphysical graviton modes depend on $\bol{k}$, and substitute the $A^{\mu\mu}$'s 
by the $a^{ii}$'s, Eq.~(\ref{eq:a6}), so that the integrand of $\big\{ Q, Q^{\dagger} \big\}$ restricted to the graviton sector
becomes
\begin{multline}
2\, \omega^2 \, \big[ +2\, {A^{03}}^{\dagger} A^{03} + 
{A^{01}}^{\dagger} A^{01} + {A^{02}}^{\dagger} A^{02} +{A^{13}}^{\dagger} A^{13}+
   {A^{23}}^{\dagger} A^{23} \big] +\\
 +\omega^2 \, \big[ +
{{a}^{11}}^{\dagger} {a}^{11}+ 
{{a}^{22}}^{\dagger} {a}^{22}+
{{a}^{33}}^{\dagger} {a}^{33} + {{a}^{11}}^{\dagger}{a}^{22}  + {{a}^{22}}^{\dagger}{a}^{11} 
+ \frac{1}{4} a^{\dagger}a    \big]\, .
\label{eq:80}
\end{multline}
With the definitions
\begin{gather}
J_{\pm}(\bol{k}):=\frac{ {{a}}^{11}(\bol{k}) \pm {{a}}^{22}(\bol{k})}{ \sqrt{2}}\, , \nonumber \\ 
\big[ J_{\pm}(\bol{k}), J_{\pm}(\bol{p})^{\dagger}\big]=\delta^{\sss (3)}(\bol{k}-\bol{p})\, , \quad 
\big[ J_{\pm}(\bol{k}), J_{\mp}(\bol{p})^{\dagger}\big]=0 \, ,
\label{eq:81}
\end{gather}
we find that the integrand of $\big\{ Q, Q^{\dagger} \big\}$ restricted to the graviton sector now reads 
\begin{multline}
2\, \omega^2 \, \big[ + {A^{01}}^{\dagger} A^{01} + {A^{02}}^{\dagger} A^{02} + 2\, {A^{03}}^{\dagger} A^{03}  
+{A^{13}}^{\dagger} A^{13}+\\
 +  {A^{23}}^{\dagger} A^{23} +  \frac{1}{8} a^{\dagger}a 
+\frac{1}{2}{{a}^{33}}^{\dagger} {a}^{33} +J_{+}^{\dagger}J_{+} \big]\, ,
\label{eq:82}
\end{multline}
which is manifestly the sum of  particle number operators for unphysical modes of the graviton field in the 
chosen reference frame: the two remaining physical modes for fixed $\bol{k}$ are created from the 
Fock vacuum $|\Omega\rangle$ by $J_{-}(\bol{k})^{\dagger}$ and $A^{12}(\bol{k})^{\dagger}$ in close analogy 
to the classical reduction of the degrees of freedom in a plane gravitational tensor wave 
$h^{\alpha\beta}_{cl.}(x)=\varepsilon_{cl.}^{\alpha\beta}(\bol{k})e^{-i\,k\cdot x}$
with polarization tensor $\varepsilon_{cl.}^{\alpha\beta}(\bol{k})$. 

Therefore Eq.~(\ref{eq:b4}) defines  the physical subspace in  a correct manner.
A different formulation of the graviton quantization can be found in~\cite{Grigo}.

In addition, we can  compute the generators of the time evolutions of the fields $H^{\alpha\beta}(x)$ and $\Phi(x)$, respectively.
These free quantum fields satisfy the Heisenberg equations of motion
\begin{equation}
-i\dot H^{\alpha\beta}(x)=\big[\mathbf{H}_{\sss H},H^{\alpha\beta}(x)\big]\, , \quad 
-i\dot \Phi (x)=\big[ \mathbf{H}_{\sss \Phi},\Phi(x)\big]\, .
\label{eq:89}
\end{equation}
The  Hamilton operators are easily found by~(\ref{eq:z1}) and read
\begin{gather}
\mathbf{H}_{\sss H}=\int\! \!d^{3}p\, \omega( \bol{p}) \bigg[+ \sum_{\nu =0}^{3} 
{A^{\nu\nu}(\bol{p})}^{\dagger}A^{\nu\nu}(\bol{p}) 
+2 \sum_{i=1}^{3}{ A^{0i}(\bol{p})}^{\dagger}A^{0i}(\bol{p}) \nonumber \\
+2 \sum_{\substack{i,j=1\\i<j}}^{3}{ A^{ij}(\bol{p})}^{\dagger} A^{ij}(\bol{p}) \bigg]\, ,\nonumber \\
\mathbf{H}_{\sss \Phi}=
\frac{1}{4}\int\! \!d^{3}p\, \omega( \bol{p})\, a(\bol{p})^{\dagger} a(\bol{p})\, .
\label{eq:90}
\end{gather}
If we restrict these operators to the physical subspace ${\mathcal{F}_{phys}}$ and, in addition, 
choose the special reference frame as before, we find that the integrand of $\big( \mathbf{H}_{\sss H}  
+\mathbf{H}_{\sss \Phi} \big)$ reads
\begin{equation}
\omega\, \Big[ J_{-}^{\dagger}J_{-} +2 {A^{12}}^{\dagger} A^{12} \Big]\, .
\label{eq:91}
\end{equation}
This expression can be recast with
\begin{gather}
a_{\pm}(\bol{p})^{\dagger} := \frac{1}{\sqrt{2}} J_{-}(\bol{p})^{\dagger}  \mp i\, A^{12}(\bol{p})^{\dagger}\, ,\nonumber \\
\big[ a_{\pm}(\bol{k}), a_{\pm}(\bol{p})^{\dagger}\big]=\delta^{\sss (3)}(\bol{k}-\bol{p})\, , \quad 
\big[ a_{\pm}(\bol{k}), a_{\mp}(\bol{p})^{\dagger}\big]=0 \, ,
\label{eq:88}
\end{gather}
into the form
\begin{equation}
\omega\, \Big[ a_{+}^{\dagger}a_{+} +a_{-}^{\dagger}a_{-} \Big] \, ,
\label{eq:92}
\end{equation}
which confirms that graviton states in ${\mathcal{F}_{phys}}$ have only two independent components,
the other eight  being unphysical. The four operators $a_{\pm}, a_{\pm}^{\dagger}$ absorb  or create  
physical states  with helicities $\pm$ 2.

\section{Unitarity}
\label{unitarity}
\setcounter{equation}{0}

The property of unitarity in QG, as in non-Abelian gauge theories, is very important (and usually very difficult to
prove), because the Fock space  contains a lot of unphysical states (see Sec.~\ref{sec:physical}).

Nevertheless, there exists a physical subspace ${\mathcal{F}_{phys}}$, such that the $S$-matrix restricted to
${\mathcal{F}_{phys}}$ is unitary. Because of
the unphysical degrees of freedom involved in the theory, unitarity does not hold on the whole Fock space.
There, the theory is pseudo-unitary, namely unitary with respect to the $\mathrm{K}$-conjugation.

Since $\big( u^{\nu}\big)^{\mathrm{K}}= u^{\nu}$, 
$ \big( \tilde{u}^{\nu}\big)^{\mathrm{K}}=- \tilde{u}^{\nu}$, ${H^{\alpha\beta}}^{\mathrm{K}}=H^{\alpha\beta}$
and $\Phi^{K}=\Phi$, it follows that $T_{1}^{h+u}$  is skew-$\mathrm{K}$-conjugate
\begin{equation}
\left(T_{1}^{h+u}(x)\right)^{\mathrm{K}}=-T_{1}^{h+u}(x)=\tilde{T}_{1}^{h+u}(x)\, ,
\label{eq:u1}
\end{equation}
and this holds for all the $n$-point distributions $T_{n}$ by induction ~\cite{Scha1,ym4} if the normalization constants
in the distribution  splitting Eq.~(\ref{eq:7}) are
chosen appropriately:
\begin{equation}
{T_{n}(X)}^{\mathrm{K}}=\tilde{T}_{n}(X)\, ,
\label{eq:u2}
\end{equation}
where $X:=\{x_{1},\ldots, x_{n}\}$ and $\tilde{T}_{n}(X)$ is the $n$-point distribution belonging to the perturbative
expansion of the inverse $S$-matrix. According to Eq.~(\ref{eq:u2}), we get \textit{pseudo-unitarity}
\begin{equation}
{S(g)}^{\mathrm{K}}=S(g)^{-1}\, .
\end{equation}
We cannot expect unitarity on the whole Fock space  because the \textit{scalar graviton}  $\Phi$, the $0i$-components
of $H^{\alpha\beta}$ and the ghosts are not hermitian (with respect to $\dagger$), but only skew-hermitian.

With \textit{unitarity on the physical subspace} we mean the heuristic equation
\begin{equation}
\lim_{g\uparrow 1}\, \Big[P_{phys}\,S(g)^{\dagger}\,P_{phys}\Big]\, \Big[P_{phys}\, S(g)\,P_{phys}\Big]=P_{phys}\, ,
\label{eq:u3}
\end{equation}
where $P_{phys}$ stands for the projection operator  onto ${\mathcal{F}}_{phys}$.
In~\cite{gri4}, we were  able to prove the perturbative version of Eq.~(\ref{eq:u3}), namely
\begin{equation}
\tilde{T}_{n}^{P}(X)=P_{phys}\,T_{n}(X)^{\dagger}\,P_{phys} +{\mathrm{divergences}}\, ,
\label{eq:u4}
\end{equation}
where $\tilde{T}_{n}^{P}(X)$ is the $n$-point distribution of the $S$-matrix inverted on $\mathcal{F}_{phys}$
\begin{multline}
\big( P_{phys}\, S(g)\, P_{phys}\big)^{-1}=P_{phys}+\sum_{n=1}^{\infty}\frac{1}{n!} \int\!\! d^{4}\,x_{1}
\ldots d^{4}x_{n}\, \tilde{T}_{n}^{P}(x_{1},\ldots,
x_{n})\\ \times g(x_{1})\cdot\ldots\cdot g(x_{n})\,.
\end{multline}
The sum of divergences appearing on the right side of Eq.~(\ref{eq:u4}) does  not harm, because the divergences  can be 
integrated out  in the adiabatic limit $g\to 1$.

\section{Scalar Matter Coupled to Quantum Gravity}
\label{sec:matter}
\setcounter{equation}{0}

In this section we investigate scalar massive matter fields coupled to QG; also the massless limit will be discussed.

Expanding $\mathcal{L}_{\sss M}=\frac{1}{2}\sqrt{-g}
\left( g^{\mu\nu}\phi_{;\mu}\phi_{;\nu}-m^2 \phi^2 \right)$ as in Sec.~\ref{sec:gravity}, we find
\begin{multline}
\mathcal{L}_{\sss M}=\underbrace{\frac{1}{2}\left( \eta^{\mu\nu}\phi_{,\mu}\phi_{,\nu}-m^2\phi^2\right)}_
{=\mathcal{L}_{\sss M}^{\sss (0)} } 
+\underbrace{\frac{\kappa}{2}\, h^{\mu\nu}\left(\phi_{,\mu}\phi_{,\nu}-\frac{m^2}{2}\eta_{\mu\nu}\phi^2\right)}_
{=\kappa \mathcal{L}_{\sss M}^{\sss (1)} } \\
+\underbrace{ \frac{m^2 \kappa^2}{8}\left( h^{\alpha\beta}h^{\alpha\beta}\phi\phi
   -\frac{1}{2} h h\phi\phi\right)}_{={\kappa^2}\mathcal{L}_{\sss M}^{\sss (2)} }
  + O(\kappa^3)\, .
\label{eq:ma}
\end{multline}
From the first term we obtain the Klein-Gordon equation of motion
\begin{equation}
(\Box + m^2)\phi(x)=0\,.
\end{equation}
We quantize the scalar field by imposing the commutation rule
\begin{equation}
\left[ \phi(x), \phi(y) \right]=-i\, D_{ m}(x-y) \, .
\label{eq:mat}
\end{equation}
The first order
matter-graviton coupling reads
\begin{equation}
T_{1}^{\sss M}(x)= i\,\kappa\, :\! \mathcal{L}_{\sss M}^{\sss (1)}(x) \! : =\frac{i\kappa}{2}\,\bigg\{ :\! 
h^{\alpha\beta} \phi_{,\alpha}\phi_{,\beta}\! : -\frac{m^2}{2}\, :\! h \phi\phi \! : \bigg\}=\frac{i\kappa}{2}\, 
:\!h^{\alpha\beta} b_{\alpha\beta\mu\nu} T^{\mu\nu}_{\sss M}\! : \, ,
\end{equation}
where $T^{\mu\nu}_{\sss M}$ is the conserved energy-momentum tensor of the matter field $T^{\mu\nu}
_{\sss M}=\phi^{,\mu}\phi^{,\nu}-\eta^{\mu\nu}\mathcal{L}_{\sss M}^{\sss (0)}$. 
Gauge invariance to first order is readily
established
\begin{equation}
d_{Q}T_{1}^{\sss M}(x)=\partial_{\mu}^{x}\,\frac{\kappa}{2}\,\bigg\{ :\! u^{\nu}\phi_{,\mu}\phi_{,\nu}\!: 
   -\frac{1}{2}:\! u^{\mu}\phi_{,\nu}\phi_{,\nu}\!: +\frac{m^2}{2}:\! u^{\mu}\phi\phi\!: \bigg\}
=\partial_{\mu}^{x}\, T^{\mu\, \sss  M}_{ 1/1}(x)\, .
\end{equation}

\subsection{Tree Graph Sector}

Gauge invariance to second order  in the tree graph sector
\begin{equation}
d_{Q}T_{2}^{tree}(x,y)=\mathrm{divergence}-\frac{\kappa^2 m^2}{2}:\! u^{\alpha}(x)_{,\beta}
h^{\alpha\beta}(x)\phi(x)\phi(x)\!:\,  \delta^{\sss (4)}(x-y)\,,
\label{eq:m1}
\end{equation}
is spoiled by the local term on the right side that cannot be written as a divergence. By exploiting the ambiguity in
the normalization, in order to restore gauge invariance we can add on both sides of Eq.~(\ref{eq:m1})
\begin{equation}
d_{Q}N_{2}^{tree}(x,y)=\frac{\kappa^2 m^2}{2}:\! u^{\alpha}_{,\beta}
h^{\alpha\beta}\phi\phi\!:\,  \delta^{\sss (4)}(x-y)\,,
\end{equation}
which is the gauge variation of the normalization term
\begin{equation}
N_{2}^{tree}(x,y)=\frac{i\kappa^2 m^2}{4}\,\left( :\! h^{\alpha\beta}h^{\alpha\beta}\phi\phi \! :
   -\frac{1}{2}:\!  h h\phi\phi\! : \right)\delta^{\sss (4)}(x-y)\,,
\end{equation}
so that we arrive at $d_{Q}T_{2}^{tree}(x,y)+d_{Q}N_{2}^{tree}(x,y)=
{divergence}$, see~\cite{gri6}. 

Also in this case, gauge invariance to second order requires the introduction of a new matter--graviton local
interaction which turns out to agree with the second order in  the classical expansion of $\mathcal{L}_{\sss M}$ 
given in Eq.~(\ref{eq:ma}), up to the factor $1/2$ coming 
from the scattering matrix.

In the massless case we obtain directly $d_{Q}T_{2}^{tree}(x,y)
={divergence}$, because the local anomaly of Eq.~(\ref{eq:m1}) does not appear. This agrees with the fact
that for massless matter $\mathcal{L}_{\sss M}^{\sss (j)}=0,\; \forall  j \ge 2 $.

\subsection{Graviton Self-Energy}
\label{sec:GSE}

As in  the calculation of  Sec.~\ref{sec:causal}, we find a contribution to the graviton self-energy
 tensor if we perform two
matter field contractions in $D_{2}(x,y)=\big[ T_{1}^{\sss M}(x),T_{1}^{\sss M}(y)\big]$
\begin{gather}
D_{2}^{\sss SE}(x,y)= :\!h^{\alpha\beta}(x)h^{\mu\nu}(y)\! :\,  
d_{2}^{\sss SE}(x-y)_{\alpha\beta\mu\nu}\, ,\nonumber \\
\hat{d}_{2}^{\sss SE}(p)_{\alpha\beta\mu\nu}=\frac{\kappa^2 \pi}{960 (2\pi)^4}\Big[
\hat{P}(p)^{\sss (4)}_{\alpha\beta\mu\nu} +\frac{m^2}{p^2}
      \hat{Q}(p)^{\sss (4)}_{\alpha\beta\mu\nu}
+\frac{m^4}{p^4}\hat{R}(p)^{\sss (4)}_{\alpha\beta\mu\nu} \Big]\nonumber \\
\times \sqrt{1-\frac{4m^2}{p^2}}\, \Theta(p^2-4m^2)\, \mathrm{sgn}(p^0)\, ,
\label{eq:m2}
\end{gather}
where the three polynomials of degree 4 have the same structure as in Eq.~(\ref{eq:21}) with the coefficients:
$\hat{P}_{\alpha\beta\mu\nu}=[-8,-4,1,-1,-1]$,
$\hat{Q}_{\alpha\beta\mu\nu}=[-16,-8,-8,8,-12]$ and
$\hat{R}_{\alpha\beta\mu\nu}=[-48,-24,+16,-16,+4]$.

We split the scalar distribution in Eq.~(\ref{eq:m2}) according to their  singular order $\omega = 0$, $\omega = -2$ and $\omega = -4$,
respectively, because the polynomials can be neglected in the splitting (\cite{gri6}),  and find
\begin{equation}
T_{2}^{\sss SE}(x,y)= :\!h^{\alpha\beta}(x)h^{\mu\nu}(y)\! :\,  
i\, \Pi(x-y)^{tot}_{\alpha\beta\mu\nu}\, .
\end{equation}
The graviton self-energy tensor is evaluated using Eq.~(\ref{eq:8}) and reads (admitting  also a freedom in the 
 normalization)
\begin{multline}
\hat{\Pi}(p)^{\sss tot}_{\alpha\beta\mu\nu}=\frac{\kappa^2 \pi}{960 (2\pi)^5}\Bigg\{ \Big[
       \hat{P}(p)^{\sss (4)}_{\alpha\beta\mu\nu} +\frac{m^2}{p^2}
       \hat{Q}(p)^{\sss (4)}_{\alpha\beta\mu\nu} +\frac{m^4}{p^4}
       \hat{R}(p)^{\sss (4)}_{\alpha\beta\mu\nu} \Big]\, \hat{\Pi}(p)\nonumber \\
 +\frac{m^2}{6p^2}\,\hat{R}(p)^{\sss (4)}_{\alpha\beta\mu\nu}
 +\sum_{a=0}^{2} \hat{N}(p)^{(2a)}_{\alpha\beta\mu\nu} \Bigg\} \, ,
\label{eq:m3}
\end{multline}
where the normalization terms $\hat{N}(p)_{(2a)}^{\alpha\beta\mu\nu}$ must be chosen in a gauge 
invariant way (see below).  The scalar distribution $\hat{\Pi}(p)$ is given by 
\begin{equation}
\begin{split}
\hat{\Pi}(p)&=\int_{q}^{\infty} \!\! ds \frac{\sqrt{s(s-q)}}{s^2(1-s+i0)} \qquad ,\,\qquad \text{where }\,\: q=\frac{4m^2}{p^2} \\
            &=-2 -\sqrt{1-q}\Big( \log \left| \frac{1-\sqrt{1-q}}{1+\sqrt{1-q}}\right| +i\pi \Big)\Theta(1-q)\\
            &\quad +2\sqrt{q-1} \arctan \left( \frac{1}{\sqrt{q-1}} \right)\Theta(q-1)\, .
\end{split}
\end{equation}
The above calculation shows that, in our approach, scalar matter  coupled to QG  does not require the introduction
of a non-renormalizable counterterm~\cite{Hooft,deser2}, {\em i.e.} of a  counterterm that cannot be absorbed in the
redefinitions of bare parameters appearing in the original Lagrangian of the theory. 

Further, gauge invariance 
\begin{gather}
d_{Q}T_{2}^{\sss SE}(x,y)=\partial_{\sigma}^{x}\big( :\!u^{\rho}(x)h^{\mu\nu}(y)\! :\,  
b^{\alpha\beta\rho\sigma} \Pi(x-y)^{ tot}_{\alpha\beta\mu\nu}\big)+(x\leftrightarrow y) \,,
\end{gather}
implies the identity
\begin{equation}
b^{\alpha\beta\rho\sigma}\,\d_{\sigma}^{x}\,\Pi(x-y)^{tot}_{\alpha\beta\mu\nu}=0\,,
\end{equation}
which corresponds to the transversality of the $2$-point connected Green function ( or \textit{Slavnov--Ward identity})
\begin{equation}
p^{\alpha}\,\hat{G}(p)_{\alpha\beta\mu\nu}^{\sss [2]}=p^{\alpha}\,
\Big[ b_{\alpha\beta\gamma\delta}\,\hat{D}_{0}^{\sss F}(p)\,
\hat{\Pi}(p)_{tot}^{\gamma\delta\rho\sigma}\,b_{\rho\sigma\mu\nu}\,\hat{D}_{0}^{\sss F}(p)\Big]=0\,.
\end{equation}
The  attached line represents  a  free graviton Feynman propagator
\begin{equation}
\langle \Omega | T \{  h^{\alpha\beta}(x) h^{\mu\nu}(y) \}  | \Omega \rangle = -i\, b^{\alpha\beta\mu\nu}
D_{0}^{\mathrm{\sss{F}}}(x-y)\,.
\end{equation}
This latter is affected by  radiative corrections  due to self-energy insertions.
If we require that the mass of the graviton and the coupling constant remain unchanged, we find for
the graviton self-energy  tensor
\begin{multline}
\hat{\Pi}(p)^{ tot}_{\alpha\beta\mu\nu}=\frac{\kappa^2 \pi}{960 (2\pi)^5}\,\Bigg\{ \Big[
       \hat{P}(p)^{\sss (4)}_{\alpha\beta\mu\nu} +\frac{m^2}{p^2}
       \hat{Q}(p)^{\sss (4)}_{\alpha\beta\mu\nu} +\frac{m^4}{p^4}
       \hat{R}(p)^{\sss (4)}_{\alpha\beta\mu\nu} \Big]\hat{\Pi}(p) \\
 +\frac{m^2}{6p^2}\,\hat{R}(p)^{\sss (4)}_{\alpha\beta\mu\nu}
+ z_{1}\,\hat{Z}_{1}(p)^{\sss (4)}_{\alpha\beta\mu\nu}
+ z_{2}\,\hat{Z}_{2}(p)^{\sss (4)}_{\alpha\beta\mu\nu} \Bigg\} \, ,
\end{multline}
where $\hat{Z}_{i}(p)^{\sss (4)}_{\alpha\beta\mu\nu},\, i=1,2$ are 2 fixed gauge invariant polynomials  and 
$z_{i}\in\mathbb{R},\, i=1,2$. 

In the massless case we obtain
\begin{equation}
\hat{\Pi}(p)^{tot}_{\alpha\beta\mu\nu}=\frac{\kappa^2 \pi}{960 (2\pi)^5}\,
         \hat{P}(p)^{\sss (4)}_{\alpha\beta\mu\nu}\, \log\left(\frac{-(p^2+i0)}{M_{0}^2}\right)\, .
\end{equation}
This massless particle loop gives also a  correction to the Newtonian potential
\begin{equation}
V(r)=\frac{-G\,m_{1}\,m_{2}}{r}\,\bigg( 1 +\frac{1}{10} \frac{G\, \hbar}{c^3\,\pi\, r^2}\bigg)\, .
\end{equation}

\subsection{Matter Self-Energy}

In $D_{2}$ we can isolate also the matter self-energy contribution by performing one graviton and one matter field
contraction, in this case we obtain
\begin{gather}
D_{2}^{\sss  MSE}(x,y)= :\! \phi(x)\phi(y) \! : d_{2}^{\sss MSE}(x-y) \, ,\nonumber \\
\hat{d}_{2}^{\sss  MSE}(p)=\frac{-\kappa^2 m^2 \pi}{2 (2\pi)^4}\left( p^2 -\frac{m^2}{2}\right)
\left( 1 -\frac{m^2}{p^2}\right) \, \Theta(p^2-m^2)\, \mathrm{sgn}(p^{0}) \, .
\end{gather}
We split the numerical distribution $\hat{d}_{2}^{\sss MSE}$ with $\omega(\hat{d}_{2}^{\sss MSE})=2$    and obtain
\begin{gather}
T_{2}^{\sss MSE}(x,y)= :\! \phi(x)\phi(y) \! :\, i\,\Sigma(x-y) \, , \nonumber\\
\hat{\Sigma}(p)=\frac{-\kappa^2 m^2 \pi}{2 (2\pi)^5}\,\Big[ \left( p^2 -\frac{3 m^2}{2}   +\frac{m^4}{2 p^2}\right)
\left\{ \log \left|\frac{p^2-m^2}{m^2}\right| -i\pi\Theta(p^2-m^2) \right\}\nonumber \\
 +\frac{m^2}{2}-\frac{5 p^2}{4}+c_{0}+c_{2} p^2  \Big] \, .
\end{gather}
If we formally sum the series of graphs with 0, 1, 2,\dots self-energy insertions, we obtain the matter propagator
\begin{equation}
\begin{split}
\hat{\Sigma}(p)^{\sss tot} & =+\hat{D}_{ m}^{\mathrm{\sss{F}}}(p)+
\hat{D}_{m}^{\mathrm{\sss{F}}}(p)\, (2\pi)^4\, \hat{\Sigma}(p) \,\hat{D}_{m}^{\mathrm{\sss{F}}}(p)+\\
&\quad + \hat{D}_{m}^{\mathrm{\sss{F}}}(p)\, (2\pi)^4\,  \hat{\Sigma}(p)\,
\hat{D}_{m}^{\mathrm{\sss{F}}}(p)\,(2\pi)^4\, \hat{\Sigma}(p)\,
\hat{D}_{m}^{\mathrm{\sss{F}}}(p)+\ldots \\
 & =- (2\pi)^{-2}\left( p^2-m^2 +i0+(2\pi)^2 \hat{\Sigma}(p)\right)^{-1}\, .
\end{split}
\end{equation}
The mass-normalization condition reads  $\hat{\Sigma}(p^2=m^2)=0$ and fixes $c_{0}=m^2(\frac{3}{4}-c_{2})$.
To find a condition for $c_{2}$ one should consider the vertex function
$\hat{\Lambda}(p,q)_{\alpha\beta}$ to the third order in the $3$-point distribution 
$T_{3}(x,y,z)=:\f(x)\f(y) h^{\alpha\beta}(z)\!:\Lambda(x,y,z)_{\alpha\beta}$
as in QED~\cite{Scha1}. 

In the massless case we obtain $D_{2}^{\sss MSE}(x,y)=T_{2}^{\sss MSE}(x,y)=0$.

\subsection{Vacuum Graphs}

If we perform three contractions in $D_{2}$, we get the vacuum graph contribution
\begin{gather}
\hat{D}_{2}^{\sss VG}(p) =\frac{\kappa^2 m^2}{(2\pi)^5}\, 
                               \Theta(p^2-4m^2)\, \mathrm{sgn}(p^0)\,\hat{f}(p)\, ,\nonumber \\
\hat{f}(p)=\frac{1}{384}\,\sqrt{1-\frac{4m^2}{p^2}}\,\Bigg(p^4-7m^2 p^2+6m^4\Bigg)+\nonumber\\
     +  \frac{m^6}{16 p^2}\,\log\left(\sqrt{\frac{p^2}{4m^2}}+\sqrt{\frac{p^2}{4m^2} -1}\right)\,.
\end{gather}
After distribution splitting with $\omega(\hat{D}_{2}^{\sss VG})=4$ we obtain
\begin{equation}
\hat{T}_{2}^{\sss VG}(p) =\hat{X}(p)^{an} +\frac{\kappa^2 m^2}{2(2\pi)^5}\, \hat{f}(p)\,
      \Theta(p^2-4m^2)\, ,
\label{eq:v1}
\end{equation}
where $\hat{X}(p)^{an}$ is the analytic continuation of $\hat{R}_{2}^{\sss VG}(p)$
\begin{multline*}
\hat{X}(p)^{an}=\frac{i\kappa^2 m^2}{384 (2\pi)^6}\,\Bigg[ -3 p^4 +\frac{31}{2}m^3p^2-6m^4+
\Big(-p^4+7m^2p^2-6m^4 \Big)\sqrt{1-\frac{4m^2}{p^2}} \\
\times \log\left(\frac{\sqrt{1-4m^2/p^2} -1}{\sqrt{1-4m^2/p^2} +1}\right)+
\frac{24m^6}{p^2}\,\log^{2}\bigg(\sqrt{\frac{-p^2}{4m^2}} +\sqrt{1-\frac{p^2}{4m^2}}\bigg)\Bigg] \, .
\end{multline*}
Since we are interested in the adiabatic limit of the vacuum graphs as in Sec.~\ref{subsec:vac}, we isolate in
Eq.~(\ref{eq:v1}) the leading behaviour in the limit $p^2\to 0$ (IR-regime)
\begin{equation}
\hat{T}_{2}^{\sss VG}(p) \sim \underbrace{ \frac{-i\kappa^2}{5120(2\pi)^6}}_{=:B}\, p^6 + O(p^8)\, ,
\end{equation}
so that the adiabatic limit  becomes in scaling form
\begin{equation}
\begin{split}
\lim_{g\to 1}\big(\Omega ,S_{2}(g)\Omega\big) & =\frac{(2\pi)^{2}}{2}\lim_{\epsilon\to 0}
\frac{1}{\epsilon ^{4}}\int\!\!d^{4}p\Big[
 \hat{T}_{2}^{\sss VG}(\epsilon p)+
\hat{N}_{2}^{\sss VG}(\epsilon p)\Big]\ \hat{g}_{0}(p)\,  \hat{g}_{0}(-p) \\
& = \frac{(2\pi)^{2}}{2}\lim_{\epsilon\to 0} \frac{1}{\epsilon ^{4}}\int\!\!d^{4}p\Big[ B\epsilon^{6} p^6 +O(p^8)
+c_{0}+\\
&\hspace{35mm}+c_{2}\epsilon^2 p^2 +c_{4}\epsilon^4 p^4 \Big]\ \hat{g}_{0}(p)\,  \hat{g}_{0}(-p) \\
& = 0\, .
\end{split}
\end{equation}
The existence of the above limit requires $c_{0}=c_{2}=0$ and is assured by the IR-behaviour of the massive theory
$\hat{T}_{2}^{\sss VG}(p)\sim p^6$ for $p^2 \to 0$. Independence from the test functions $g_{0}$ is
reached by choosing  $c_{4}=0$.

\section{Abelian Gauge Fields Coupled to Quantum Gravity}\label{sec:maxwell}

We discuss very briefly the coupling between gravitons and $U(1)$-Abelian gauge fields (\textit{photons}), see~\cite{gri5} for 
the details.

We expand the Lagrangian $\mathcal{L}_{\sss A}=-\sqrt{-g} F_{\mu\nu}F_{\alpha\beta}g^{\alpha\mu}
g^{\beta\nu}/ 4 $ in powers of the coupling constant $\kappa$ and isolate the first order coupling
\begin{equation}
\begin{split}
T_{1}^{\sss A}(x)& =i\,:\!\mathcal{L}_{\sss A}^{\sss (1)}(x)\!:
=i\, \frac{\kappa}{2}\, :\! h^{\alpha\beta}(x) T_{\sss A}(x)_{\alpha\beta}\! : \\
& = i\, \frac{\kappa}{2}\, :\! h^{\alpha\beta}\big(-F_{\beta\nu}F_{\alpha}^{\ \nu}+\frac{\eta_{\alpha\beta}}{4}F_{\mu\nu}
F^{\mu\nu}\big)\!:
\end{split}
\end{equation}
where $F^{\mu\nu}=\partial^{\mu}A^{\nu}-\partial^{\nu}A^{\mu}$.
The photon field is quantized according to
\begin{equation}
\left[ A^{\mu}(x),A^{\nu}(y)\right]=i\, \eta^{\mu\nu}\, D_{0}(x-y)\, .
\end{equation}

First order gauge invariance,
 $d_{Q}T_{1}^{\sss EM}(x)={divergence}$, holds true because of 
$T_{{\sss A}\ ,\nu }^{\mu\nu}=0$ and $\eta_{\alpha\beta}T_{{\sss A}}^{\alpha\beta}=0$.

We evaluate some loop contributions~\cite{capduff,Deser,Deservan} in second order of perturbation theory.
The photon loop graviton self-energy  contribution~\cite{capduff}  reads
\begin{equation}
T_{2}^{\sss SE}(x,y)= :\!h^{\alpha\beta}(x)h^{\mu\nu}(y)\! :\,  
i\, \Pi(x-y)_{\alpha\beta\mu\nu}\, ,
\end{equation}
where the self-energy tensor reads
\begin{equation}
\begin{split}
\hat{\Pi}(p)^{\alpha\beta\mu\nu}=\frac{\kappa ^2 \pi}{960 (2\pi)^5}\,
\bigg[
&-16 \,  p^{\alpha}p^{\beta}p^{\mu}p^{\nu}
 -8  \,  p^2\big( p^{\alpha}p^{\beta}\eta^{\mu\nu}+p^{\mu}p^{\nu}\eta^{\alpha\beta}\big)\\
&+12 \,  p^2 \big(p^{\alpha}p^{\mu}\eta^{\beta\nu}+p^{\alpha}p^{\nu}\eta^{\beta\mu}+p^{\beta}p^{\mu}\eta^{\alpha\nu}+
                   p^{\beta}p^{\nu}\eta^{\alpha\mu}\big) \\
-12 \,  p^4 \big(\eta^{\alpha\mu}\eta^{\beta\nu}&+\eta^{\alpha\nu}\eta^{\beta\mu}\big)
+8  \,  p^4 \eta^{\alpha\beta}\eta^{\mu\nu}\bigg]\, \log\left(\frac{-(p^2+i0)}{M^2}\right) 
\end{split}
\end{equation}
and satisfies the perturbative gauge invariance condition (and as a consequence the Slavnov--Ward identity) 
as in Sec.~(\ref{sec:GSE}), and, in addition, is transversal  $p_{\alpha}\hat{\Pi}(p)^{\alpha\beta\mu\nu}=0$ and traceless  
$\eta_{\alpha\beta}\hat{\Pi}(p)^{\alpha\beta\mu\nu}=0$. 

The photon loop graviton self-energy contributes also  to the corrections of the Newtonian  potential as the graviton and ghost loop:
\begin{equation}
V(r)=\frac{-Gm_{1}m_{2}}{r}\,\bigg( 1 +\frac{4}{15}\, \frac{G\, \hbar}{c^3\,\pi\, r^2}\bigg)\, .
\end{equation}

The photon self-energy contribution through a graviton-photon loop reads
\begin{equation}
T_{2}^{\sss  SE}(x,y)= :\!A^{\gamma}(x)A^{\rho}(y)\! :\,\big ( -i\, \Pi(x-y)_{\gamma\rho}\big)\, ,
\end{equation}
where the photon self-energy tensor is
\begin{equation}
\hat{\Pi}^{\gamma\rho}(p)=\frac{\kappa^2 \pi}{12(2\pi)^5}\,\Big(p^{2}\,p^{\gamma}p^{\rho} -\eta^{\gamma\rho}\,p^{4}\Big)\,
\log\left(\frac{-(p^2+i0)}{M^2}\right)\, .
\end{equation}
In both cases, we find UV-finite and cutoff-free  results for our one-loop calculations. Therefore the introduction of counterterms 
(that cannot be renormalized away, see~\cite{deser2,Deser,Deservan}) is not necessary.

\section{General Ansatz for Matter Coupling and Perturbative  Gauge Invariance}\label{sec:general}

In this section we adopt a new strategy~\cite{Aste1,Schawell} in order to construct a gauge invariant
theory of quantum gravity coupled to matter fields. 

This purely quantum approach relies merely on the inductive
causal construction of $T_{n}$ (see Sec.~\ref{sec:smatrix}) and on the perturbative quantum gauge invariance condition
(see  Sec.~\ref{sec:gauge}). It does not  appeal to any classical Lagrangian density  and uses only free quantum fields.
In~\cite{Schawell} this idea was implemented for  pure QG, as already explained above.

\subsection{Massive Case}

We adopt the same strategy by choosing the following ansatz for the most general massive matter coupling (disregarding
non-relevant divergence couplings) between one  graviton and two  matter fields
\begin{equation}
T_{1}^{\sss M}(x)=i\, \frac{\kappa}{2}\,\big(+ \tilde{x} :\!h^{\alpha\beta}\phi_{,\alpha}\phi_{,\beta}\! : 
+y :\!h\phi_{,\gamma}\phi_{,\gamma}\! : +z :\!h^{\alpha\beta}\phi\phi_{,\alpha\beta}\! :
+w\, m^2 :\!h\phi\phi\! :\big) \, ,
\label{eq:gen1}
\end{equation}
where $\tilde{x},y,z,w \in\mathbb{R}$ are undetermined coefficients. 
The quantized graviton and matter fields satisfy the commutation rules Eq.~(\ref{eq:12}) and Eq.~(\ref{eq:mat}),
respectively.

The condition of perturbative gauge invariance to first order, $d_{Q}T_{1}^{\sss M}={divergence}$,
implies $y=z/2 - w-\tilde{x}/2$.

In second order, for the graviton self-energy $T_{2}^{\sss SE}$, gauge
invariance $d_{Q}T_{2}^{\sss SE}={divergence}$ (which is equivalent to the Slavnov--Ward identity)
implies $w=-\tilde{x}/{2}$ so that the general matter coupling becomes
\begin{equation}
T_{1}^{\sss M}(x)=i\, \frac{\kappa}{2}\big(+ \tilde{x} :\!h^{\alpha\beta}\phi_{,\alpha}\phi_{,\beta}\! : 
+\frac{z}{2} :\!h\phi_{,\gamma}\phi_{,\gamma}\! : +z :\!h^{\alpha\beta}\phi\phi_{,\alpha\beta}\! :
-\frac{\tilde{x}}{2}\, m^2 :\!h\phi\phi\! :\big) \, ,
\end{equation}
with only two  undetermined coefficients instead of four. 

The analysis of perturbative gauge invariance to second order
in the tree sector gives two possible solutions:
\begin{equation}
d_{Q}T_{2}^{\sss  tree}(x_{1},x_{2})+d_{Q}N_{2}^{\sss tree}(x_{1},x_{2})={divergence}\quad \Longleftrightarrow
\ \tilde{x}=z\quad \mathrm{or} \quad \tilde{x}=z+1\, .
\end{equation}
Both of the conditions on the right side are in agreement with the natural assumption that the first order coupling
can be written as
\begin{equation}
T_{1}^{\sss M}(x)=i\, \frac{\kappa}{2}\, :\!h^{\alpha\beta}(x)\,b_{\alpha\beta\mu\nu}\,
   \Theta^{\mu\nu}_{\sss M}(x)\! :
\end{equation}
for an \textit{improved} energy-momentum tensor with $\Theta^{\mu\nu}_{\sss M}(x)_{,\nu}=0$.
The $b^{\alpha\beta\mu\nu}$-tensor appears here because we are using the expansion of the Goldberg variable.

Thus, we have seen that if we start with the most general ansatz for $T_{1}^{\sss M}$, Eq.~(\ref{eq:gen1}), 
with four  undetermined parameters, then perturbative gauge invariance up to the second order is able to reduce 
this number to one. Analysis  of the third order should then fix unambiguously this last parametric freedom.

\subsection{Massless Case}

As in the previous section, we investigate if the condition of perturbative quantum gauge invariance 
is strong enough to select by itself, among all  the possible couplings between massless matter fields and gravitons,  
the \textit{right} coupling, namely to select only one coupling which, in addition, should agree with the expansion of
the classical Lagrangian.

Let us write the most general ansatz for the massless matter coupling (disregarding unimportant divergence couplings) as
\begin{equation}
\begin{split}
T_{1}^{\sss M}(x)& =i\, \frac{\kappa}{2}\,\big(+ \tilde{x} :\!h^{\alpha\beta}\phi_{,\alpha}\phi_{,\beta}\! : 
+y :\!h\phi_{,\gamma}\phi_{\gamma}\! : +z :\!h^{\alpha\beta}\phi\phi_{,\alpha\beta}\! :\big)  \\
&=i\, \frac{\kappa}{2}\, :\!h^{\alpha\beta}(x)\,b_{\alpha\beta\mu\nu}\,
   \Theta^{\mu\nu}_{\sss M}(x)\! :\, ,
\label{eq:gen2}
\end{split}
\end{equation}
where $\tilde{x},y,z\in\mathbb{R}$ are undetermined coefficients and $\Theta^{\mu\nu}_{\sss M}$ an
\textit{improved} energy-momentum tensor with $\Theta^{\mu\nu}_{\sss M}(x)_{,\nu}=0$. 

To establish a connection between our undetermined coefficients $\tilde{x},y,z$ and the classical theory, we expand 
the non-minimally matter coupled Lagrangian 
\begin{equation}
\tilde{\mathcal{L}}_{\sss M}=\frac{1}{2}\,\sqrt{-g}\,\big(g^{\mu\nu}\phi_{;\mu}\phi_{;\nu} +\xi\, R\, \phi^2\big)
\label{eq:gen3}
\end{equation}
in terms of the graviton field and compare the coefficients. We obtain the relations: $\tilde{x}=1-2\xi$, $y=\xi$ and $ z=-2\xi$.

If $\xi=1/6$ we obtain the \textit{Callan-Jackiw improved} energy-momentum tensor~\cite{callan}.

On the other side we can also consider in Eq.~(\ref{eq:gen2}) the most general conserved and traceless 
energy-momentum tensor
\begin{equation}
\Theta^{\mu\nu}_{\sss M}=\alpha\, \phi^{,\mu}\phi^{,\nu}-\frac{\alpha}{4}\,\eta^{\mu\nu}\,\phi_{,\gamma}
\phi^{,\gamma}-\frac{\alpha}{2}\,\phi\phi^{,\mu\nu}\ , \quad \alpha\in \mathbb{R}\, ,
\label{eq:gen4}
\end{equation}
which gives the relations  $\tilde{x}=\alpha$, $y=-\alpha /4$ and $z=-\alpha /2 $.

Gauge invariance to first order, $d_{Q}T_{1}^{\sss M}={divergence}$ is then always satisfied.

Gauge invariance to second order for the matter loop graviton self-energy , $d_{Q}T_{2}^{\sss SE}={divergence}$,
requires that $y=z/2$. 

Since the particle circulating in the loop is massless, we expect the 
self-energy  tensor to be traceless, too. This implies $y=-\tilde{x}/4$ and $z=-\tilde{x}/2$.

With these relations among the parameters we can undertake the investigation of perturbative 
gauge invariance to second order in the tree graph sector. We find again
\begin{equation}
d_{Q}T_{2}^{\sss tree}(x_{1},x_{2})+d_{Q}N_{2}^{\sss tree}(x_{1},x_{2})={divergence}\quad \Longleftrightarrow
\ \tilde{x}=z\quad \mathrm{or} \quad \tilde{x}=z+1\, .
\end{equation}
Obviously the first relation $\tilde{x}=z$ cannot be satisfied by our coefficients in both cases, Eq.~(\ref{eq:gen3}) and
Eq.~(\ref{eq:gen4}), therefore should be rejected.

The second relation $\tilde{x}=z+1$ is satisfied $\forall \xi \in \mathbb{R}$ in the case of non-minimal matter 
coupling $\xi R \phi^2$. The reason is that this term has zero gauge variation so that its addition to the
term $:\!h^{\alpha\beta}\phi_{,\alpha}\phi_{,\beta}\! :$, which is already gauge invariant
to first and second order alone, does not change the theory from the point of view of the gauge structure.

On the other side, if we examine  the relation $\tilde{x}=z+1$ in view of Eq.~(\ref{eq:gen4}), we find that it has
only one solution, namely $\alpha=2/3$. Therefore, according to our strategy, perturbative quantum gauge invariance
to first and second order, together with some assumptions  about  the structure of the massless matter energy-momentum
tensor, Eq.~(\ref{eq:gen4}), leads to the coupling
\begin{equation}
T_{1}^{\sss M}(x)=i\, \frac{\kappa}{2}\,\big(
+ \frac{2}{3}\, :\!h^{\alpha\beta}\phi_{,\alpha}\phi_{,\beta}\! : 
-\frac{1}{6}\, :\!h\phi_{,\gamma}\phi_{,\gamma}\! : -\frac{1}{3}\, :\!h^{\alpha\beta}\phi\phi_{,\alpha\beta}\! :\big)\, .
\end{equation}
This result is equivalent to the choice of $\xi=1/6$ in Eq.~(\ref{eq:gen3}) and, equivalently, to the use of the 
\textit{Callan-Jackiw improved} energy-momentum tensor.

\end{document}